\documentclass[12pt]{article}

\usepackage[mathscr]{eucal}
\usepackage{amsfonts}
\usepackage{amssymb}
\usepackage{amsthm}

\newcommand{\be}{\begin{equation}}
\newcommand{\ee}{\end{equation}}
\newcommand{\bea}{\begin{eqnarray}}
\newcommand{\eea}{\end{eqnarray}}

\def\theequation{\arabic{section}.\arabic{equation}}
\begin{document}

\begin{titlepage}
\vspace*{0.5cm}

\renewcommand{\thefootnote}{$\ddagger$}
\begin{center} {\LARGE\bf  New spinorial particle model in tensorial }

\vspace{0.5cm}

{\LARGE\bf  space-time and interacting higher spin fields}

\vspace{2cm}

{\large\bf Sergey Fedoruk},${}^{\dagger}$\footnote{\,\,On leave of absence from V.N.\,Karazin Kharkov National University, Ukraine}\qquad{\large\bf Jerzy
Lukierski} ${}^{\ast}$

\vspace{1.5cm}

${}^{\dagger}${\it Bogoliubov  Laboratory of Theoretical Physics, JINR,}\\ {\it Joliot-Curie 6, 141980 Dubna, Moscow region, Russia} \\ \vspace{0.1cm}

{\tt fedoruk@theor.jinr.ru}\\ \vspace{0.5cm}

${}^{\ast}${\it Institute for Theoretical Physics, University of Wroc{\l}aw,}\\ {\it pl. Maxa Borna 9, 50-204 Wroc{\l}aw, Poland} \\ \vspace{0.1cm}

{\tt lukier@ift.uni.wroc.pl}\\

\vspace{2cm}



\end{center} \vspace{0.2cm} 

\begin{abstract} \noindent The Maxwell-covariant particle model is formulated in tensorial extended $D=4$ space-time $(x_\mu,z_{\mu\nu})$ parametrized by
ten-dimensional coset of $D=4$ Maxwell group, with added auxiliary Weyl spinors $\lambda_\alpha$, $y^{\alpha}$. We provide the Hamiltonian quantization of
the model and demonstrate that first class constraints modify the known equations obtained for massless higher spin fields in flat tensorial space-time. We
obtain the Maxwell-covariant field equations for new infinite dimensional spin multiplets. The component fields assigned to different spin values are
linked by couplings proportional to rescaled electromagnetic coupling constant $\tilde e=e\,m$, where $m$ is the mass-like parameter introduced in our
model. We discuss briefly the geometry of our tensorial space-time with constant torsion and its relation with the presence of constant electromagnetic
background. \end{abstract}



\newpage

\end{titlepage}

\setcounter{footnote}{0}

\setcounter{equation}{0} \section{Introduction}

The development in higher spin (HS) theories shows the importance of dynamics in generalized spaces with supplemented additional tensorial coordinates
 (see e.g. \cite{BL,BLS,BLPS,Vas,PST,BBAST,Vas1}).
In particular, in $D=4$ the massless free HS fields can be derived from the quantization of the spinorial particle model on flat tensorial space, described
by $D=4$ Minkowski space extended by six tensorial coordinates \cite{BL,BLS,Vas} generated by the tensorial central charges. We would like to point out
that six commuting with each other tensorial charges are supplemented as well if we enlarge the Poincare algebra to the Maxwell algebra \cite{Bac,Schr,B,Sor,BG,BGKL}. The
corresponding Maxwell tensorial space-time, generated by fourmomenta and six additional tensorial charges, is endowed in tensorial sector with constant
torsion proportional to  electromagnetic coupling constant $e$. The aim of this paper is to consider the spinorial particle model in ten-dimensional $D=4$
tensorial space-time with torsion described by the coset of $D=4$ Maxwell group. It follows that the additional tensorial coordinates can be linked with
the spin degrees of freedom, and the model after first quantization will describe linear equations for coupled infinite-component HS field multiplets
(Maxwell HS fields).

The Maxwell algebra \cite{Bac,Schr} is obtained as the following enlargement of the Poincare algebra with generators ($P_{\alpha\dot\beta}$,
$M_{\alpha\beta}$, $\bar M_{\dot\alpha\dot\beta}$) by six commuting with each other tensorial charges $Z_{\alpha\beta}=Z_{\beta\alpha}$, $\bar
Z_{\dot\alpha\dot\beta}=(Z_{\alpha\beta})^+$ \footnote{We shall consider in this paper the case $D=4$ and use two-spinor notation, i.e.
$P_{\alpha\dot\beta}=\sigma^{\mu}_{\alpha\dot\beta}P_{\mu}$, $Z_{\alpha\beta}=\sigma^{\mu\nu}_{\alpha\beta}Z_{\mu\nu}$, $\bar
Z_{\dot\alpha\dot\beta}=\tilde\sigma^{\mu\nu}_{\dot\alpha\dot\beta}Z_{\mu\nu}$, where
$(\sigma^\mu)_{\alpha\dot\alpha}=(1_2,\vec{\sigma})_{\alpha\dot\alpha}$,
$(\tilde\sigma^\mu)^{\dot\alpha\alpha}=\epsilon^{\alpha\beta}\epsilon^{\dot\alpha\dot\beta}(\sigma^\mu)_{\beta\dot\beta}=
(1_2,-\vec{\sigma})^{\dot\alpha\alpha}$, $\sigma^{\mu\nu}=i\,\sigma^{[\mu}\tilde\sigma^{\nu]}$, $\tilde\sigma^{\mu\nu}=i\,\tilde\sigma^{[\mu}\sigma^{\nu]}$
$\sigma^{\mu\nu}_{\alpha\beta}=\epsilon_{\beta\gamma}(\sigma^{\mu\nu})_{\alpha}{}^{\gamma}$,
$\tilde\sigma^{\mu\nu}_{\dot\alpha\dot\beta}=\epsilon_{\dot\beta\dot\gamma}(\tilde\sigma^{\mu\nu})^{\dot\gamma}{}_{\dot\alpha}$. Always in this paper we
use weight coefficient in (anti)symmetrization, i.e. $A_{(\alpha}B_{\beta)}=\frac12\,(A_{\alpha}B_{\beta}+A_{\beta}B_{\alpha})$,
$A_{[\alpha}B_{\beta]}=\frac12\,(A_{\alpha}B_{\beta}-A_{\beta}B_{\alpha})$. } \footnote{In (\ref{M-alg}) we do not introduce any dimensionfull parameter,
i.e. we assume that the mass dimension of $Z_{\alpha\beta}$, $\bar Z_{\dot\alpha\dot\beta}$ is 2 (twice of mass dimension of $P_{\alpha\dot\beta}$). We
stress that one can introduce any mass dimensionality $[Z_{\alpha\beta}]=[\bar Z_{\dot\alpha\dot\beta}]$ of tensorial generators by introducing suitable
dimensionfull constant in (\ref{M-alg}). If we assume $[Z_{\alpha\beta}]=[\bar Z_{\dot\alpha\dot\beta}]=1$ we get the case of tensorial coordinates with
dimensionalities as in \cite{BL,BLS,BLPS,Vas,PST,BBAST,Vas1}; if $[Z_{\alpha\beta}]=[\bar Z_{\dot\alpha\dot\beta}]=0$ such form of Maxwell algebra was
considered in \cite{AKLuk,Sor2,Durka}. In Sect.\,1-3 below we shall use only the parameterless form  (\ref{M-alg}-\ref{Lor-alg}) of Maxwell algebra. The
consequences of introducing various dimensionfull parameters in (\ref{M-alg}) for the basic equations derived in this paper we shall consider in Appendix.}
\begin{equation}\label{M-alg} \begin{array}{rrc} &&[P_{\alpha\dot\alpha},P_{\beta\dot\beta}] = 2 i\,e \left(\epsilon_{\dot\alpha\dot\beta}Z_{\alpha\beta}+
\epsilon_{\alpha\beta}\bar Z_{\dot\alpha\dot\beta}\right),\\[7pt] &&[Z_{\alpha\beta},Z_{\gamma\delta}] = [Z_{\alpha\beta},\bar Z_{\dot\alpha\dot\beta}]=
[Z_{\alpha\beta},P_{\gamma\dot\gamma}]=0 \,, \end{array} \end{equation} where \begin{equation}\label{M-alg1} \!\!\!\!\!\!\!\!\!\!\!\!\begin{array}{rrc}
&&[M_{\alpha\beta},P_{\gamma\dot\gamma}] = -i\epsilon_{\gamma(\alpha}P_{\beta)\dot\gamma}\,,\qquad [\bar M_{\dot\alpha\dot\beta},P_{\gamma\dot\gamma}] =
-i\epsilon_{\dot\gamma(\dot\alpha}P_{\gamma\dot\beta)}\,, \\[7pt] &&[M_{\alpha\beta},Z_{\gamma\delta}] = i\left( \epsilon_{\alpha\gamma}Z_{\beta\delta}+
\epsilon_{\beta\delta}Z_{\alpha\gamma}\right), \quad [\bar M_{\dot\alpha\dot\beta},\bar Z_{\dot\gamma\dot\delta}] = i\left(
\epsilon_{\dot\alpha\dot\gamma}\bar Z_{\dot\beta\dot\delta}+ \epsilon_{\dot\beta\dot\delta}\bar Z_{\dot\alpha\dot\gamma}\right), \quad
[M_{\alpha\beta},\bar Z_{\dot\gamma\dot\delta}] = 0 \end{array} \end{equation} and $M_{\alpha\beta}=M_{\beta\alpha}$, $\bar
M_{\dot\alpha\dot\beta}=(M_{\alpha\beta})^+$ describe the Lorentz algebra generators \begin{equation}\label{Lor-alg}
\!\!\!\!\!\!\!\!\!\!\!\!\!\!\!\!\!\begin{array}{rrc} &&[M_{\alpha\beta},M_{\gamma\delta}] = i\left( \epsilon_{\alpha\gamma}M_{\beta\delta}+
\epsilon_{\beta\delta}M_{\alpha\gamma}\right), \,\,\,\, [\bar M_{\dot\alpha\dot\beta},\bar Z_{\dot\gamma\dot\delta}] = i\left(
\epsilon_{\dot\alpha\dot\gamma}\bar M_{\dot\beta\dot\delta}+ \epsilon_{\dot\beta\dot\delta}\bar M_{\dot\alpha\dot\gamma}\right), \,\,\,\,
[M_{\alpha\beta},\bar M_{\dot\gamma\dot\delta}] = 0. \end{array} \end{equation} The quantity $e$ in  (\ref{M-alg}) is the dimensionless electromagnetic
coupling constant.

One can introduce Maurer-Cartan (MC) one-forms ($\omega_{(P)}^{\alpha\dot\beta}$, $\omega_{(Z)}^{\alpha\beta}$, $\omega_{(\bar Z)}^{\dot\alpha\dot\beta}$)
on ten-dimensional coset ($\cal{M}$ is the Maxwell group) which defines $D=4$ proper Maxwell group ${\cal{M}}_0$ \begin{equation}\label{M-coset}
{\cal{M}}_0=\frac{\cal{M}}{O(3,1)}=e^{i(z^{\alpha\beta}Z_{\alpha\beta}+\bar z^{\dot\alpha\dot\beta}\bar Z_{\dot\alpha\dot\beta})}
e^{ix^{\alpha\dot\beta}P_{\alpha\dot\beta}}\,. \end{equation} By analogy with the constant torsion in $N=1$ Wess-Zumino superspace, the tensorial Maxwell
space-time described by proper Maxwell group (\ref{M-coset}) is endowed with constant torsion.

The simplest Maxwell generalization of standard relativistic $D=4$ particle model was considered in \cite{FedLuk}, where it was extended to ten-dimensional
tensorial space-time manifold (\ref{M-coset}). It was shown in \cite{FedLuk} that after first quantization such a model presents in Lorentz-covariant way
the $D=4$ particle interacting with electromagnetic (EM) field characterized by a constant field strength (so-called Landau orbit problem). In this paper
we shall generalize the $D=4$ spinorial particle model defined on flat tensorial space-time with supplemented Weyl spinor variable introduced in
\cite{BL,BLS,Vas}. Let us recall that the spinorial massless $D=4$ $N=1$ superparticle model was firstly  described by the following SUSY-invariant action,
proposed by Shirafuji \cite{Shir} \begin{equation}\label{Shir-act} S^{N=1\,SUSY}=\int \,
\lambda_{\alpha}\bar\lambda_{\dot\beta}\,\omega_{(\mathscr{P})}^{\alpha\dot\beta}\,, \end{equation} where the one-form
$\omega_{(\mathscr{P})}^{\alpha\dot\beta}$ is derived from the superalgebra $\{Q_{\alpha},\bar Q_{\dot\beta}\}=2\mathscr{P}_{\alpha\dot\beta}$ and
$\lambda_{\alpha}$, $\bar\lambda_{\dot\alpha}=(\lambda_{\alpha})^+$ is an auxiliary two-component Weyl spinor. Below we shall consider the Maxwell
counterpart of Shirafuji model, which by observing the correspondence $\Big(Q_\alpha, \bar Q_{\dot\beta}; \mathscr{P}_{\alpha\dot\beta} \Big)$
$\leftrightarrow$ $\Big(P_{\alpha\dot\beta}; Z_{\alpha\beta}, \bar Z_{\dot\alpha\dot\beta} \Big)$ and $\omega_{(\mathscr{P})}^{\alpha\dot\beta}$
$\leftrightarrow$ $\Big(\omega_{(Z)}^{\alpha\beta}, \bar\omega_{(\bar Z)}^{\dot\alpha\dot\beta} \Big)$ we define as follows
\begin{equation}\label{Max-act-m} S^{Max} =\int \, \Big( a\lambda_{\alpha}\lambda_{\beta}\omega_{(Z)}^{\alpha\beta} +\bar
a\bar\lambda_{\dot\alpha}\bar\lambda_{\dot\beta}\bar\omega_{(\bar Z)}^{\dot\alpha\dot\beta} \Big)\,. \end{equation}

Because the tensorial coordinates $(z^{\alpha\beta},\bar z^{\dot\alpha\dot\beta})$ and one forms $(\omega_{(Z)}^{\alpha\beta},\bar\omega_{(\bar
Z)}^{\dot\alpha\dot\beta})$ have mass dimensionality equal to $-2$ ($[z^{\alpha\beta}]=[\bar z^{\dot\alpha\dot\beta}]=
[\omega_{(Z)}^{\alpha\beta}]=[\bar\omega_{(\bar Z)}^{\dot\alpha\dot\beta}]=-2$; see footnote 2) if as usual we assign
$[\lambda^{\alpha}]=[\bar\lambda^{\dot\alpha}]=\frac12$, we obtain that $[a]=1$. By fixing global $U(1)$ phase transformations
$\lambda_{\alpha}=e^{i\varphi}\lambda_{\alpha}$, $\bar\lambda_{\dot\alpha}=e^{-i\varphi}\bar\lambda_{\dot\alpha}$, which commute with Lorentz
$SL(2;\mathbb{C})$ transformations, the complex parameter $a$ can be made real ($a=\bar a=m$) and introduces in the model a mass-like parameter $m$.

It will be shown (see Sect.\,3) that $m$ is not providing standard notion of mass for particular HS field components; its presence will be seen in the
terms describing couplings between $D=4$ HS fields with different spins. Further, in order to obtain in our model nontrivial ``conformal limit'' $m\to 0$
we shall add to the action (\ref{Max-act-m}) the action (\ref{Shir-act}) with $\mathscr{P}_\mu$ (commuting Poincare momenta) replaced however by the
noncommuting Maxwell momenta $P_\mu$ from (\ref{M-alg}).

In order to perform effectively the quantization which provides the HS field equations in our Maxwell tensorial space $X
=(x^{\alpha\dot\alpha},z^{\alpha\beta}, \bar z^{\dot\alpha\dot\beta} )$ we shall also add to the action  (\ref{Max-act-m}) the free kinetic term linear in
time derivatives of $\lambda_{\alpha}$, $\bar\lambda_{\dot\alpha}$.\footnote{ Such terms in generalized Shirafuji model in $D=4$ tensorial space obtained
by adding to Poincare algebra the tensorial central charges was proposed by Vasiliev \cite{Vas}. }

We recall that the $D=4$ massless conformal fields were obtained as describing first quantization of particle model in flat tensorial space  $X^A
=(x^{\alpha\dot\alpha},z^{\alpha\beta}, \bar z^{\dot\alpha\dot\beta} )$, with additional two-tensor coordinates $z^{\mu\nu}$=($z^{\alpha\beta}$, $\bar
z^{\dot\alpha\dot\beta}$) generated by the tensorial central charges appearing in generalized $D=4$ Poincare superalgebra.\footnote{ Firstly such tensorial
supercharges were postulated in $D=11$ superalgebra what led to the notion of M-algebra \cite{Cur,AzTown,Sezg}. } Further there was considered the
derivation of HS fields in $D=4$ AdS by quantization of the particle model on non-flat tensorial superspace described by the group manifold $Sp(4)$
\cite{BLPS,PST,BBAST}.
In AdS space-time there appears a geometric dimensionfull parameter, AdS radius or cosmological constant, which permits
to introduce interacting higher spin fields for spins $s>2$ \cite{FV87a,FV87b}.
In this paper we introduce other modification of flat tensorial space, characterized by alternative way
of introducing the dimensionfull parameter. In the formulation of Maxwell algebra (\ref{M-alg})
without geometric dimensionfull parameter (see footnote 2)) the mass-like parameter $m$
is dynamical, appears in the action (see (\ref{Maxwell-Lagr})). However one
can change the mass dimensionality of the generators $Z_{\alpha\beta}$, $\bar Z_{\dot\alpha\dot\beta}$ and dual tensorial coordinates
$z^{\alpha\beta}$, $\bar z^{\dot\alpha\dot\beta}$ if we introduce in the relation (\ref{M-alg}) a suitable geometric dimensionfull parameter. In general case one can
replace (\ref{M-alg}) as follows
\begin{equation}\label{M-alg-m}
[P_{\alpha\dot\alpha},P_{\beta\dot\beta}] = 2 i\,eM^{\,\xi}
\left(\epsilon_{\dot\alpha\dot\beta}Z_{\alpha\beta}+ \epsilon_{\alpha\beta}\bar Z_{\dot\alpha\dot\beta}\right)\,,
\end{equation}
where $\xi$ is a real
number, $M$ describes a geometric mass parameter ($[M]=1$), $e$ is dimensionless ($[e]=0$). The value $\xi=0$ was introduced in original Maxwell algebra (\ref{M-alg})
with fourmomenta commutator proportional to dimensionless electromagnetic coupling constant.
If $\xi=1$  the geometric mass-like parameter $M$ enters into the Maxwell algebra,
and one can show that in the dynamical equations of the corresponding particle model the geometric parameter $M$
replaced the dynamical parameter $m$ (see Appendix).
In agreement with the discussion of spin two barrier for higher spin interactions
(see e.g. \cite{BekBS}) we conjecture that it is the presence of new dimensionfull
parameter which permits our framework with coupled higher spin fields.
Because in our model with Maxwell symmetries the particle action contains the
mass parameter $m$,  it implies that the Maxwell-invariant HS dynamics is nonconformal. We shall present below such dynamics, what should help to arrive
at the physical interpretation of additional tensorial coordinates $z^{\alpha\beta}$, $\bar z^{\dot\alpha\dot\beta}$ parametrizing the Maxwell group manifold.

In Sect.\,2 we shall perform the canonical quantization of the model (\ref{Max-act-m}) with supplemented kinetic term for $\lambda^{\alpha}$,
$\bar\lambda^{\dot\alpha}$.\footnote{ The additional kinetic term linear in time derivatives, more proper for spinorial degrees of freedom, was used in
\cite{FZ,Vas,FI}. } By using the phase space formulation we shall specify the set of first and second class constraints. It appears that in first quantized
theory the first class constraints will describe the set of field equations for new higher spin multiplets in the tensorial space $X
=(x^{\alpha\dot\alpha},z^{\alpha\beta}, \bar z^{\dot\alpha\dot\beta} )$ which define new HS Maxwell dynamics. Such equations will describe the
generalization of the known ``unfolded equations'' \cite{BLS,Vas,PST,Vas1} for massless HS free fields with flat space-time derivatives
$\partial_{\alpha\dot\beta}$ replaced by the Maxwell-covariant derivatives $D_{\alpha\dot\beta}$. Important property of our particle model is that the
Casimirs of Maxwell algebra \cite{Schr,Sor} \begin{eqnarray}\label{Max-Cas1} C^{Max}_1 &=& P_{\alpha\dot\beta}P^{\alpha\dot\beta} +4e\Big(M_{\alpha\beta}
Z^{\alpha\beta} +\bar M_{\dot\alpha\dot\beta}\bar Z^{\dot\alpha\dot\beta} \Big)\,, \\[3pt] \label{Max-Cas23} C^{Max}_2 &=& Z_{\alpha\beta} Z^{\alpha\beta}
\,,\qquad\qquad C^{Max}_3 =\bar Z_{\dot\alpha\dot\beta}\bar Z^{\dot\alpha\dot\beta}\,, \\[3pt] \label{Max-Cas4} C^{Max}_4 &=& 2Z^{\alpha\beta}\bar
Z^{\dot\alpha\dot\beta}P_{\alpha\dot\alpha}P_{\beta\dot\beta}  - {\textstyle\frac12}\left(Z^{\gamma\delta}Z_{\gamma\delta}+ \bar
Z^{\dot\gamma\dot\delta}\bar Z_{\dot\gamma\dot\delta}\right) P^{\alpha\dot\alpha}P_{\alpha\dot\alpha}\\ &&\qquad\qquad\qquad\quad\,\,\, +\,
2e\left(Z^{\gamma\delta}Z_{\gamma\delta}- \bar Z^{\dot\gamma\dot\delta}\bar Z_{\dot\gamma\dot\delta}\right) \left(Z^{\alpha\beta}M_{\alpha\beta}- \bar
Z^{\dot\alpha\dot\beta}\bar M_{\dot\alpha\dot\beta}\right)\nonumber \end{eqnarray} will vanish as a consequence of first class constraints. It follows from
(\ref{Max-Cas1}) that for vanishing value of $C^{Max}_1$ the formula for $D=4$ mass square $M^2=\frac12\,P_{\alpha\dot\beta}P^{\alpha\dot\beta}$ is linear
in Lorentz generators describing relativistic angular momenta, what suggests some link with the known mass formulae for Regge trajectories.\footnote{ The
idea of linking $D=2$ Maxwell algebra with Regge trajectories in two-dimensional stringy field theory was firstly suggested by D.\,Soroka and V.\,Soroka
\cite{SS-D}. }

In Sect.\,3 we shall describe the new linear set of field equations for space-time fields describing the infinite-dimensional nonconformal Maxwell-HS
multiplets. Similarly like in SUSY-covariant field theory one uses superspace and the covariant odd derivatives $D_{\alpha}$, $\bar D_{\dot\alpha}$, the
Maxwell-covariant formulation is given by extended tensorial space-time $X=(x_\mu, z_{\mu\nu})$ and Maxwell-covariant space-time derivatives $D_\mu$. If we
expand the fields on Maxwell tensorial space into $D=4$ HS fields with arbitrary Lorentz spins, in comparison with known equations for decoupled free
massless conformal HS fields we obtain the equations with new space-time-dependent terms, which link fields with different values of the Lorentz spins
$(j_1,j_2)$. We shall also show that the second order equations, generalizing the Klein-Gordon equation for the corresponding scalar Maxwell-HS fields can
be described by the bilinear Casimir (\ref{Max-Cas1}) with Maxwell algebra generators $(P_{\alpha\dot\beta},M_{\alpha\beta}, \bar M_{\dot\alpha\dot\beta},
Z_{\alpha\beta},\bar Z_{\dot\alpha\dot\beta} )$ suitably realized in ten-dimensional tensorial space  $X$ describing the generalization of $D=4$
space-time.

We recall that at present there was considered the quantization of spinorial particle model of Shirafuji type on two $D=4$ tensorial manifolds: the flat
one, described by $R^{10}$ \cite{BL,BLS,PST}, and described by the group manifold $Sp(4)$ \cite{BLS,BLPS,Vas,PST,BBAST,Vas1}. Because both tensorial
manifolds can be described by the same coset $Sp(8)/[GL(4)\,{\subset\!\!\!\!\!\!\times} K_{10}]$ ($K_{10}$ is ten-dimensional Abelian group of generalized
conformal translations) \cite{PST}, they provide two different choices of coordinates on the same manifold, and consequently corresponding free particle
models have the same massless spectrum of free conformal HS particles. At present it is not clear if there is a way of ``diagonalizing'' the interacting
Maxwell HS multiplets, and the question of their mass spectrum is the problem for further investigation. Subsequently  at the end of Sect.\,3 we show that
the torsion of Maxwell space-time can be interpreted as describing the coupling to Abelian gauge potential. 
Finally in Appendix we shall consider a general scale reparametrization of the
basic algebraic relation (\ref{M-alg}), modifying  the ``canonical'' dimensionality $[Z_{\alpha\beta}]=[\bar Z_{\dot\alpha\dot\beta}]=2$ and introducing
additional, more geometric mass-like parameter $M$.

\setcounter{equation}{0} \section{Maxwell-covariant spinorial particle model and \\ its formulation in generalized phase space}

\subsection{Covariant Maurer-Cartan one-forms}

Taking exponential parametrization (\ref{M-coset}) and using algebraic relations (\ref{M-alg}), (\ref{M-alg1}) we obtain \begin{equation}\label{om-def}
{\cal M}_0^{-1}d {\cal M}_0=i\Big(\omega^{\alpha\dot\beta}P_{\alpha\dot\beta}+\omega^{\alpha\beta}Z_{\alpha\beta} + \bar\omega^{\dot\alpha\dot\beta}\bar Z
_{\dot\alpha\dot\beta} \Big)\,, \end{equation} where the Maurer-Cartan one-forms are \begin{equation}\label{M-om} \begin{array}{ccc}
&&\omega^{\alpha\dot\beta}= d x^{\alpha\dot\beta}\,,\\[5pt] &&\omega^{\alpha\beta}=  d z^{\alpha\beta} + e\,x^{(\alpha\dot\gamma}d
x^{\beta)}_{\dot\gamma}\,,\qquad \bar\omega^{\dot\alpha\dot\beta}= d\bar z^{\dot\alpha\dot\beta}+ e\,x^{\gamma(\dot\alpha}d x^{\dot\beta)}_{\gamma}\,.
\end{array} \end{equation} The dual Maxwell-covariant derivatives $D_A$=($D_{\alpha\dot\beta}$, $D_{\alpha\dot\beta}$, $\bar D_{\dot\alpha\dot\beta}$) are
given by the formulae \begin{equation}\label{M-der} \begin{array}{ccc} && D_{\alpha\dot\beta}= -i\left(\displaystyle\frac{\partial}{\partial
x^{\alpha\dot\beta}} \,\,+ e\,x^{\gamma}_{\dot\beta}\,\displaystyle\frac{\partial}{\partial z^{\alpha\gamma}}\,\,+
e\,x^{\dot\gamma}_{\alpha}\,\displaystyle\frac{\partial}{\partial \bar z^{\dot\beta\dot\gamma}}\right),\\[9pt] && D_{\alpha\beta}=
-i\,\displaystyle\frac{\partial}{\partial z^{\alpha\beta}}\,,\qquad \bar D_{\dot\alpha\dot\beta}= -i\,\displaystyle\frac{\partial}{\partial \bar
z^{\dot\alpha\dot\beta}}\,\,. \end{array} \end{equation} The relations (\ref{M-der}) defines by means of the formula $\left[D_A,D_B\right]=T_{A,B}{}^C D_C$
the nonvanishing torsion in Maxwell tensorial space-time $(x_\mu, z_{\mu\nu})$ \begin{equation}\label{tor}
T_{\mu,\,\nu}{}^{\lambda\rho}=\delta_\mu^{[\lambda} \delta_\nu^{\rho]}\,. \end{equation} The one-forms (\ref{M-om}) and vector fields (\ref{M-der}) are
invariant under the space-time (parameters $a^{\alpha\dot\beta}$) and tensorial Maxwell translations (parameters $b^{\alpha\beta}$, $\bar
b^{\dot\alpha\dot\beta}$) \begin{equation}\label{M-trans} \begin{array}{ccc} &&\delta x^{\alpha\dot\beta}= a^{\alpha\dot\beta}\,,\\[5pt] &&\delta
z^{\alpha\beta}=   b^{\alpha\beta} + x^{(\alpha\dot\gamma}a^{\beta)}_{\dot\gamma}\,,\qquad \delta\bar z^{\dot\alpha\dot\beta}= \bar
b^{\dot\alpha\dot\beta}+ x^{\gamma(\dot\alpha}a^{\dot\beta)}_{\gamma} \end{array} \end{equation} and are covariant under the Lorentz transformations
(parameters $\ell^{\alpha\beta}$, $\bar\ell^{\dot\alpha\dot\beta}$) \begin{equation}\label{M-trans-L} \delta x^{\alpha\dot\beta}=
\ell^{\alpha\gamma}x_{\gamma}^{\dot\beta}+\bar\ell^{\dot\alpha\dot\gamma}x_{\dot\gamma}^{\beta}\,,\qquad \delta z^{\alpha\beta}=
2\ell^{\alpha\gamma}z_{\gamma}^{\beta} \,,\qquad \delta\bar z^{\dot\alpha\dot\beta}= 2\bar\ell^{\dot\alpha\dot\beta}\bar z_{\dot\gamma}^{\dot\beta}\,.
\end{equation}

\subsection{Particle action and constraints}

We shall consider the following Maxwell-invariant particle action
\begin{equation}\label{Maxwell-act}
S=\int \,
\left[\lambda_{\alpha}\bar\lambda_{\dot\beta}\,\omega^{\alpha\dot\beta}+ m\left(\lambda_{\alpha}\lambda_{\beta}\,\omega^{\alpha\beta} +
\bar\lambda_{\dot\alpha}\bar\lambda_{\dot\beta}\,\bar\omega^{\dot\alpha\dot\beta}\right) \right]\,.
\end{equation}
{}From action (\ref{Maxwell-act})
follows a complicated structure of the constrains with four first class constraints. Similarly as in the previous HS particle models in tensorial
space-time \cite{BL,BLS}, the appearance of the second class constraints $p_\lambda^{\alpha}\approx0$, $\bar p_\lambda^{\dot\alpha}\approx0$ makes the
quantization difficult. It is useful to convert these constraints into the first class constraints what is achieved by adding four additional degrees of
freedom and new four gauge symmetries \cite{BLS} . Effectively, as shown in \cite{Vas}, such conversion is produced by adding to the action
(\ref{Maxwell-act}) the term with additional coordinates $(y^{\alpha},{\bar y}^{\dot\alpha})$
\begin{equation}\label{Maxwell-act1} S_\lambda=\int d\tau\,
\Big(\lambda_{\alpha}\dot y^{\alpha} + {\bar\lambda}_{\dot\alpha}\dot{\bar y}^{\dot\alpha} \Big)\,. \end{equation} As a result, the constraints
$p_\lambda^{\alpha}\approx0$, $\bar p_\lambda^{\dot\alpha}\approx0$ do not appear and  $y^{\alpha}$, $\bar y^{\dot\alpha}$ play the role of canonical
variables conjugate to $\lambda_{\alpha}$, ${\bar\lambda}_{\dot\alpha}$.

Thus, we consider in this paper the model defined by the Lagrangian
\begin{equation}\label{Maxwell-Lagr}
L=\lambda_{\alpha}\bar\lambda_{\dot\beta}\,\dot
x^{\alpha\dot\beta}+ m\lambda_{\alpha}\lambda_{\beta}\,\Big(\dot z^{\alpha\beta} + e\,x^{\alpha\dot\gamma}\dot x^{\beta}_{\dot\gamma}\Big) +
m\bar\lambda_{\dot\alpha}\bar\lambda_{\dot\beta}\,\Big( \dot{\bar z}^{\dot\alpha\dot\beta}+ e\,x^{\gamma\dot\alpha}\dot x^{\dot\beta}_{\gamma} \Big)+
\lambda_{\alpha}\dot y^{\alpha} + {\bar\lambda}_{\dot\alpha}\dot{\bar y}^{\dot\alpha}\,, \end{equation} which describes the particle motion in generalized
coordinate space $(x^{\alpha\dot\beta},z^{\alpha\beta},{\bar z}^{\dot\alpha\dot\beta}, y^{\alpha},\bar y^{\dot\alpha})$. The definitions of corresponding
momenta $(p_{\alpha\dot\beta},f_{\alpha\beta},{\bar f}_{\dot\alpha\dot\beta}, \lambda_{\alpha},\bar\lambda_{\dot\alpha})$ lead to the following constraints
in the model \footnote{ We omit below the consideration of the constraints $p_\lambda^{\alpha}-y^{\alpha}\approx0$, $\bar p_\lambda^{\dot\alpha} -\bar
y^{\dot\alpha}\approx0$ and $p_{y\,\alpha}\approx0$, $\bar p_{y\,\dot\alpha}\approx0$. After introducing for them Dirac brackets the variables
$(\lambda_{\alpha},\bar\lambda_{\dot\alpha})$ become the momenta of $(y^{\alpha},\bar y^{\dot\alpha})$. The tensorial part of the phase space
$(x^{\alpha\dot\beta},z^{\alpha\beta},{\bar z}^{\dot\alpha\dot\beta}, p_{\alpha\dot\beta},f_{\alpha\beta},{\bar f}_{\dot\alpha\dot\beta})$ is supplemented
by auxiliary spinorial phase space $(y^{\alpha},\bar y^{\dot\alpha}, \lambda_{\alpha},\bar\lambda_{\dot\alpha})$ with the canonical Poisson brackets
$\{y^\alpha, \lambda_{\beta}\}_{{}_P}=\delta_{\beta}^\alpha$, $\{\bar y^{\dot\alpha},\bar\lambda_{\dot\beta} \}_{{}_P}=\delta_{\dot\beta}^{\dot\alpha}$.
After quantization these two parts of generalized phase space commute.} \begin{eqnarray}\label{costr-x} T_{\alpha\dot\beta}&\equiv&p_{\alpha\dot\beta}+
e\,f_{\alpha\gamma}x^\gamma_{\dot\beta}+e\,\bar f_{\dot\beta\dot\gamma}x^{\dot\gamma}_{\alpha}-\lambda_{\alpha}\bar\lambda_{\dot\beta}\approx 0\, ,\\[5pt]
T_{\alpha\beta}&\equiv&f_{\alpha\beta}-m\lambda_{\alpha}\lambda_{\beta}\approx 0\,,\label{costr-y}\\[5pt] \bar T_{\dot\alpha\dot\beta}&\equiv&\bar
f_{\dot\alpha\dot\beta}-m\bar\lambda_{\dot\alpha}\bar\lambda_{\dot\beta}\approx 0\,.\label{costr-by} \end{eqnarray} It is easy to see that after insertion
of (\ref{costr-y}), (\ref{costr-by}) in (\ref{costr-x}) we obtain the covariantization of the constraints leading to unfolded equations for HS fields
\cite{BLS,Vas} \begin{equation}\label{costr-n} T_{\alpha\dot\beta}= D_{\alpha\dot\beta}-\lambda_{\alpha}\bar\lambda_{\dot\beta}\,, \end{equation} where
\begin{equation}\label{D-cov-cl} D_{\alpha\dot\beta}=p_{\alpha\dot\beta} +e\,x^\gamma_{\dot\beta}p_{\alpha\gamma}+ e\,x^{\dot\gamma}_{\alpha}\bar
p_{\dot\beta\dot\gamma}\,, \end{equation} is the classical counterpart of the Maxwell-covariant derivative.

The only nonvanishing Poisson brackets (PB) of the constraints (\ref{costr-x})-(\ref{costr-by}) are \begin{equation}\label{costr-PB}
\left\{T_{\alpha\dot\alpha}, T_{\beta\dot\beta}\right\}_{{}_P}= 2 e\left(\epsilon_{\dot\alpha\dot\beta} f_{\alpha\beta}+ \epsilon_{\alpha\beta}\bar
f_{\dot\alpha\dot\beta}\right) \approx 2e\, m \left(\epsilon_{\dot\alpha\dot\beta} \lambda_{\alpha}\lambda_{\beta}+
\epsilon_{\alpha\beta}\bar\lambda_{\dot\alpha}\bar\lambda_{\dot\beta}\right). \end{equation} Therefore, the constraints (\ref{costr-y}), (\ref{costr-by})
are first class and imply that the tensorial coordinates $(z^{\alpha\beta}, \bar z^{\dot\alpha\dot\beta})$ are becoming pure gauge degrees of freedom
whereas the constraints (\ref{costr-x}) are the superposition of two first class and two second class constraints.

We stress that such structure is not present in the particle model describing the standard free massless HS particles \cite{BLS,Vas}, where the
counterparts of the constraints  (\ref{costr-x}) are first class. Only in limit $e\to 0$ the model  (\ref{Maxwell-Lagr}) yield the standard action
\cite{BLS,Vas} of higher spin particle.

For extracting second class constraints from (\ref{costr-x}) we shall use second Weyl spinor $u_{\alpha}$, as was proposed in \cite{BLS}. Such auxiliary
spinor satisfies the condition $\lambda^\alpha u_{\alpha}=1$ and has nonvanishing PB $\{u_{\alpha},y^\beta \}_{{}_P}=u_{\alpha}u^\beta$ (see details in
\cite{BLS}). Then, considering PB of the projections \begin{equation}\label{costr-pr} T_{\lambda\bar\lambda}\equiv \lambda^\alpha
T_{\alpha\dot\alpha}\bar\lambda^{\dot\alpha}\,, \qquad T_{\lambda\bar u}\equiv \lambda^\alpha T_{\alpha\dot\alpha}\bar u^{\dot\alpha}\,, \qquad
T_{u\bar\lambda}\equiv u^\alpha T_{\alpha\dot\alpha}\bar\lambda^{\dot\alpha}\,, \qquad T_{u\bar u}\equiv u^\alpha T_{\alpha\dot\alpha}\bar
u^{\dot\alpha}\,, \end{equation} we obtain that the unique nonvanishing PB following from (\ref{costr-PB}) is \begin{equation}\label{PB-pr}
\{T_{\lambda\bar u}+T_{u\bar\lambda}, T_{u\bar u}\}_{{}_P}=4e\,m\,. \end{equation} Therefore the constraints ($T_{\lambda\bar u}+T_{u\bar\lambda}$,
$T_{u\bar u}$) are second class constraints, whereas ($T_{\lambda\bar u}-T_{u\bar\lambda}$, $T_{\lambda\bar\lambda}$) are first class.

We introduce now the conversion of the pair of second class constraints into third first class constraint by considering the equation $T_{u\bar u}\approx0$
as gauge fixing condition for the constraint $T_{\lambda\bar u}+T_{u\bar\lambda}\approx0$ generating new gauge degree of freedom. Then, we shall consider
further the constraints (\ref{costr-x}) as described by three first class constraints $T_{\lambda\bar u}\approx0$, $T_{u\bar\lambda}\approx0$,
$T_{\lambda\bar\lambda}\approx0$. Let us observe, however, that these constraints are equivalent to the projections of the constraints  (\ref{costr-x}) on
the Weyl spinor components $\lambda^\alpha$, $\bar\lambda^{\dot\alpha}$. Thus, the equivalent system, which we will quantize, is described by the phase
space variables with nonvanishing PB \begin{equation}\label{PB-c1}
\{x^{\alpha\dot\alpha},p_{\beta\dot\beta}\}_{{}_P}=\delta^{\alpha}_{\beta}\delta^{\dot\alpha}_{\dot\beta},\qquad
\{z^{\alpha\beta},f_{\gamma\delta}\}_{{}_P}=\delta^{(\alpha}_{\gamma}\delta^{\beta)}_{\delta},\qquad \{\bar z^{\dot\alpha\dot\beta},\bar
f_{\dot\gamma\dot\delta}\}_{{}_P}=\delta^{(\dot\alpha}_{\dot\gamma}\delta^{\dot\beta)}_{\dot\delta}, \end{equation} \begin{equation}\label{PB-c2}
\{y^\alpha, \lambda_{\beta}\}_{{}_P}=\delta_{\beta}^\alpha, \qquad \{\bar y^{\dot\alpha},\bar\lambda_{\dot\beta} \}_{{}_P}=\delta_{\dot\beta}^{\dot\alpha}
\end{equation} and the following first class constraints \begin{eqnarray}\label{costr-x1}
S_{\alpha}&\equiv&T_{\alpha\dot\beta}\bar\lambda^{\dot\beta}=\left( p_{\alpha\dot\beta}+ e\,f_{\alpha\gamma}x^{\gamma}_{\dot\beta}
\right)\bar\lambda^{\dot\beta} \approx D_{\alpha\dot\beta}\bar\lambda^{\dot\beta}\approx 0\, ,\\[5pt] \label{costr-x1a} \bar
S_{\dot\alpha}&\equiv&\lambda^{\beta}T_{\beta\dot\alpha}=\lambda^{\beta}\left( p_{\beta\dot\alpha}+ e\,\bar f_{\dot\alpha\dot\gamma}
x_{\beta}^{\dot\gamma}\right) \approx \lambda^{\beta} D_{\beta\dot\alpha}\approx 0\, ,\\[5pt]
T_{\alpha\beta}&\equiv&f_{\alpha\beta}-m\lambda_{\alpha}\lambda_{\beta}\approx 0\,,\label{costr-y1}\\[5pt] \bar T_{\dot\alpha\dot\beta}&\equiv&\bar
f_{\dot\alpha\dot\beta}-m\bar\lambda_{\dot\alpha}\bar\lambda_{\dot\beta}\approx 0\,.\label{costr-by1} \end{eqnarray} Because \begin{equation}
\lambda^{\alpha}S_{\alpha}\approx\bar\lambda^{\dot\alpha}\bar S_{\dot\alpha} \end{equation} the four relations (\ref{costr-x1}), (\ref{costr-x1a}) describe
only three independent first class constraints.

It is useful to make some comment about projections in (\ref{costr-x1}), (\ref{costr-x1a}). By performing these projections we omit the contribution in
first class constraints which does not depend on spinor variables and describes the field equation for spin zero case. Such contribution is present in the
following quadratic first class constraint \begin{equation}\label{constr-x3} T\equiv T_{\alpha\dot\beta}T^{\alpha\dot\beta}\approx 0\,. \end{equation}
Indeed, using $\lambda^\alpha u_\alpha=\bar\lambda^{\dot\alpha} \bar u_{\dot\alpha}=1$ we obtain $T=T_{\lambda\bar\lambda}T_{u\bar u}-T_{\lambda\bar
u}T_{u\bar\lambda}\approx 0$ because after conversion the constraints $T_{\lambda\bar u}\approx0$, $T_{u\bar\lambda}\approx0$,
$T_{\lambda\bar\lambda}\approx0$ are of first class. We add that the constraint (\ref{constr-x3}) will provide the Maxwell extension of massless
Klein-Gordon (KG) equation satisfied by the free conformal HS fields.

\subsection{Noether charges and the Casimirs}

In order to interprete the role of the first class constraints (\ref{costr-x1})-(\ref{costr-by1}), (\ref{constr-x3}) in our model we will find the Noether
currents, generated by the Maxwell generators $(P_{\alpha\dot\beta},Z_{\alpha\beta},\bar Z_{\dot\alpha\dot\beta},M_{\alpha\beta},\bar
M_{\dot\alpha\dot\beta})$ in generalized coordinate space $(x^{\alpha\dot\beta},z^{\alpha\beta},\bar z^{\dot\alpha\dot\beta}, y^{\alpha},\bar
y^{\dot\alpha})$. Using the transformations (\ref{M-trans}), (\ref{M-trans-L}) and~\footnote{The spinors $\lambda_\alpha$, $y^{\alpha}$ are inert with
respect to space-time and Maxwell translations.} \begin{equation}\label{M-trans-L1} \delta y^{\alpha}=  \ell^{\alpha\beta}y_{\beta} \,,\quad \delta\bar
y^{\dot\alpha}= \bar\ell^{\dot\alpha\dot\beta}\bar y_{\dot\beta}\,,\qquad \delta \lambda_{\alpha}=  -\ell_{\alpha\beta}\lambda^{\beta} \,,\quad \delta\bar
\lambda_{\dot\alpha}= -\bar\ell_{\dot\alpha\dot\beta}\bar\lambda^{\dot\beta} \end{equation} we obtain the following dynamical phase space realization of
Maxwell algebra generators \begin{equation}\label{M-gen-cl} \begin{array}{rcl} P_{\alpha\dot\beta}&=&-p_{\alpha\dot\beta}+e\,x_{\alpha}^{\dot\gamma}\bar
f_{\dot\gamma\dot\beta}+e\,f_{\alpha\gamma} x_{\dot\beta}^{\gamma}\,,\\[5pt] Z_{\alpha\beta}&=&-f_{\alpha\beta}\,,\qquad \qquad \bar
Z_{\dot\alpha\dot\beta}\,\,\,=\,\,\,-\bar f_{\dot\alpha\dot\beta}\,,\\[5pt] M_{\alpha\beta}&=&x_{(\alpha}^{\dot\gamma}p_{\beta)\dot\gamma} +
2z_{(\alpha}^{\gamma}f_{\beta)\gamma} +y_{(\alpha}\lambda_{\beta)}\,,\\[5pt] \bar M_{\dot\alpha\dot\beta}&=&x_{(\dot\alpha}^{\gamma}p_{\gamma\dot\beta)} +
2\bar z_{(\dot\alpha}^{\dot\gamma}\bar f_{\dot\beta)\dot\gamma} +\bar y_{(\dot\alpha}\bar \lambda_{\dot\beta)}\,. \end{array} \end{equation}

Using the expressions (\ref{M-gen-cl}) we find that the Casimirs (\ref{Max-Cas1}), (\ref{Max-Cas23}), (\ref{Max-Cas4}) of Maxwell algebra are expressed as
follows in terms of the first class constraints
 (\ref{costr-x1})-(\ref{costr-y1}), (\ref{constr-x3}):
\begin{equation}\label{Cas-fix} \begin{array}{rcl} C^{Max}_1 &\approx&T_{\alpha\dot\beta} T^{\alpha\dot\beta}+2\lambda^\alpha S_\alpha\,,\\[5pt] C^{Max}_2
&\approx&T_{\alpha\beta} T^{\alpha\beta} \,,\qquad\qquad C^{Max}_3 \approx\bar T_{\dot\alpha\dot\beta}\bar T^{\dot\alpha\dot\beta}\,,\\[5pt] C^{Max}_4
&\approx&2\lambda^\alpha \bar \lambda^{\dot\alpha} S_\alpha \bar S_{\dot\alpha}\,. \end{array} \end{equation} The numerical eigenvalues of Casimirs
(\ref{Cas-fix}) characterize the choice of infinite-dimensional Maxwell-HS irreducible field multiplets. It appears from the quantization of our
Maxwell-Shirafuji model that we obtain the Maxwell-HS realizations corresponding to all four eigenvalues of Casimirs (\ref{Cas-fix}) equal to zero.

\setcounter{equation}{0} \section{First quantization of the particle model and  interacting HS fields}

\subsection{HS field equations from first class constraints}

We consider the Schr\"{o}dinger representation in which the wave function depends on the generalized coordinates \begin{equation}\label{WF-a} \Phi
=\Phi(x^{\alpha\dot\beta},z^{\alpha\beta}, \bar z^{\dot\alpha\dot\beta}, y^\alpha,\bar y^{\dot\alpha}) \end{equation} and the generalized quantized momenta
are realized by the partial derivatives \begin{equation}\label{op-real1-a} p_{\alpha\dot\beta}=-i\frac{\partial\,\,}{\partial x^{\alpha\dot\beta}}\equiv
-i\partial_{\alpha\dot\beta}\,,\qquad f_{\alpha\beta}=-i\frac{\partial\,\,}{\partial z^{\alpha\beta}}\equiv -i\partial_{\alpha\beta}\,,\qquad \bar
f_{\dot\alpha\dot\beta}=-i\frac{\partial\,\,}{\partial \bar z^{\dot\alpha\dot\beta}}\equiv -i\bar\partial_{\dot\alpha\dot\beta}\,, \end{equation}
\begin{equation}\label{op-real2-a} \lambda_{\alpha}=-i\frac{\partial\,\,}{\partial y^{\alpha}}\equiv  -i\partial_{\alpha}\,,\qquad
\bar\lambda_{\dot\alpha}=-i\frac{\partial\,\,}{\partial \bar y^{\dot\alpha}}\equiv  -i\bar\partial_{\dot\alpha}\,. \end{equation} The Maxwell-covariant
momenta $D_{\alpha\dot\beta}$ (see (\ref{D-cov-cl})) after quantization satisfy the relation \begin{equation}\label{d-costr-com-a}
\left[D_{\alpha\dot\alpha}, D_{\beta\dot\beta}\right]= 2i\,e \left(\epsilon_{\dot\alpha\dot\beta} f_{\alpha\beta}+ \epsilon_{\alpha\beta}\bar
f_{\dot\alpha\dot\beta}\right) \end{equation} and in Schr\"{o}dinger realization describe the Maxwell-covariant derivative. Physical field equations are
defined by the quantum counterpart of first class constraints  (\ref{costr-x1})-(\ref{costr-y1}): \begin{eqnarray}\label{q-costr-x1-a} i
D_{\alpha\dot\beta}\bar\partial^{\dot\beta}\,\Phi&=&\left(\partial_{\alpha\dot\beta}+
e\,\partial_{\alpha\gamma}\,x_{\dot\beta}^{\gamma}\right)\bar\partial^{\dot\beta}\,\Phi= 0\, ,\\[5pt] \label{q-costr-z1-a} i
D_{\beta\dot\alpha}\partial^{\beta}\,\Phi&=&\left(\partial_{\beta\dot\alpha}+ e\,\bar \partial_{\dot\alpha\dot\gamma} \,x_{\beta}^{\dot\gamma}
\right)\partial^{\beta}\, \Phi= 0\, , \end{eqnarray} \begin{equation}\label{q-costr-y1-a} \Big(\partial_{\alpha\beta}+ i m\partial_{\alpha}
\partial_{\beta}\Big)\Phi=0\,,\qquad\qquad \Big(\bar \partial_{\dot\alpha\dot\beta}+ im \bar\partial_{\dot\alpha}\bar\partial_{\dot\beta}\Big)\Phi=0\,.
\end{equation}

Let us use the Taylor expansion of the wave function $\Phi$ with respect to the variables $z\equiv z^{\alpha\beta}$ and $\bar z\equiv \bar
z^{\dot\alpha\dot\beta}$. Then the constraints (\ref{q-costr-y1-a}) for the wave function \begin{equation}\label{WF-1-a} \Phi(x,z, \bar z, y,\bar y)=
\sum_{k,n=0}^{\infty}\frac{1}{k!n!}\,z^{\alpha_1\beta_1}...z^{\alpha_k\beta_k}\, \bar z^{\dot\alpha_1\dot\beta_1}...\bar z^{\dot\alpha_n\dot\beta_n}
\,\Phi^{(2k,2n)}_{\alpha_1\beta_1...\alpha_k\beta_k\,\dot\alpha_1\dot\beta_1...\dot\alpha_n\dot\beta_n}(x,y,\bar y) \end{equation} provide the expression
of all components $\Phi^{(k,n)}$, $k>0$, $n>0$ by $\Phi^{(0,0)}$ as follows: \begin{equation}\label{WF-1-comp-a}
\Phi^{(2k,2n)}_{\alpha_1\beta_1...\alpha_k\beta_k\,\dot\alpha_1\dot\beta_1...\dot\alpha_n\dot\beta_n}(x,y,\bar y)= (-im)^{k+n}\,\partial_{\alpha_1}
\partial_{\beta_1}\,...\, \partial_{\alpha_k} \partial_{\beta_k}\,\, \bar\partial_{\dot\alpha_1}\bar\partial_{\dot\beta_1}\,...\,
\bar\partial_{\dot\alpha_n}\bar\partial_{\dot\beta_n}\, \,\Phi^{(0,0)}(x,y,\bar y)\,. \end{equation} The formulae (\ref{WF-1-comp-a}) can be written down
as the solutions of eqs.\,(\ref{q-costr-y1-a}) by one compact formula \begin{equation}\label{WF-1-comp1} \Phi(x,z, \bar z, y,\bar
y)=e^{-im\left(z^{\alpha\beta}\partial_{\alpha} \partial_{\beta}+\bar z^{\dot\alpha\dot\beta}\bar\partial_{\dot\alpha}\bar\partial_{\dot\beta}\right)}
\Phi^{(0,0)}(x, y,\bar y)\,, \end{equation} what confirms the auxiliary gauge nature of the tensorial coordinates $z^{\mu\nu}=(z^{\alpha\beta},\bar
z^{\dot\alpha\dot\beta})$.

Let us analyze the remaining equations (\ref{q-costr-x1-a}), (\ref{q-costr-z1-a}) for the unique unconstrained component $\Phi^{(0,0)}(x, y,\bar y)$ of the
wave function  (\ref{WF-1-a}). We perform the following subsequent Taylor expansion \begin{equation}\label{WF-0-a} \Phi^{(0,0)}(x, y,\bar y)=
\sum_{k,n=0}^{\infty} \frac{1}{k!n!}\, y^{\alpha_1}...y^{\alpha_k}\, \bar y^{\dot\beta_1}...\bar y^{\dot\beta_n}
\,\phi^{(k,n)}_{\alpha_1...\alpha_k\,\dot\beta_1...\dot\beta_n}(x)\,, \end{equation} where
$\phi^{(k,n)}_{\alpha_1...\alpha_k\,\dot\beta_1...\dot\beta_n}(x)$ are the $D=4$ space-time Maxwell-HS fields.

Note, that in limit $e\to 0 $ the equations (\ref{q-costr-x1-a}), (\ref{q-costr-z1-a}) yield massless Dirac-Pauli-Fierz equations for massless conformal HS
fields with arbitrary helicity
\begin{equation}\label{Dir-m0-a}
e=0\,:\qquad\qquad
\partial^{\alpha_1\dot\gamma}\,\phi^{(k,n)}_{\alpha_1...\alpha_k\,\dot\beta_1...\dot\beta_n}(x)=0\,,\qquad
\partial^{\gamma\dot\beta_1}\,\phi^{(k,n)}_{\alpha_1...\alpha_k\,\dot\beta_1...\dot\beta_n}(x)=0\,.
\end{equation}
If $e\neq 0 $ we obtain the
generalization of these equations.

Now we list the equations for Maxwell-HS fields which are the consequence of (\ref{q-costr-x1-a}), (\ref{q-costr-z1-a}): \begin{itemize} \item Using
two-spinor identities $\partial^{\alpha}\partial_{\alpha}=\bar\partial^{\dot\alpha}\bar\partial_{\dot\alpha}=0$ we obtain the equation
\begin{equation}\label{Tr-Psi-a} \partial^{\alpha\dot\alpha}\partial_{\alpha}\bar\partial_{\dot\alpha}\,\Phi=0 \end{equation} which gives the generalized
Lorentz divergence conditions for the component fields in (\ref{WF-0-a}) \begin{equation}\label{Tr-comp-a}
\partial^{\alpha_1\dot\beta_1}\,\phi^{(k,n)}_{\alpha_1...\alpha_k\,\dot\beta_1...\dot\beta_n}(x)=0\,,\qquad k,n\geq 1\,. \end{equation} For $n=k=1$ we
obtain the standard Lorentz condition for the four-vector field. \item We obtain also the equations \begin{equation}\label{Eq-Psi-a}
\partial_{\dot\alpha}^{\beta}\,\partial_{\alpha}\partial_{\beta}\,\Phi=
\partial^{\dot\beta}_{\alpha}\,\bar\partial_{\dot\alpha}\bar\partial_{\dot\beta}\,\Phi \end{equation} which lead to the following set of equations
\begin{equation}\label{Eq-comp-a} \partial_{\dot\beta_{n-1}}^{\,\alpha_k}\,\phi^{(k,n-2)}_{\alpha_1...\alpha_{k-1}\alpha_k\,\dot\beta_1...\dot\beta_{n-2}}=
\partial^{\,\dot\beta_n}_{\alpha_{k-1}}\,\phi^{(k-2,n)}_{\alpha_1...\alpha_{k-2}\,\dot\beta_1...\dot\beta_{n-1}\dot\beta_n} \,,\qquad k,n\geq 2\,.
\end{equation} for the component fields. In particular for antisymmetric 2-tensors described by Lorentz spins $(2,0)+(0,2)$ we have the following equation
\begin{equation}\label{Eq-comp-M-a} \partial_{\dot\alpha}^{\,\beta}\,\phi^{(2,0)}_{\alpha\beta}=
\partial^{\,\dot\beta}_{\alpha}\,\phi^{(0,2)}_{\dot\alpha\dot\beta}\,. \end{equation} \item After inserting of (\ref{WF-0-a}) the equations
(\ref{q-costr-x1-a}), (\ref{q-costr-z1-a}) produce the following equations for component fields \begin{equation}\label{Eq-Dir-comp1-a}
\partial_{\dot\beta_{n+1}}^{\,\alpha_k}\,\phi^{(k,n)}_{\alpha_1...\alpha_k\,\dot\beta_1...\dot\beta_{n}}=
iem\,x^{\,\alpha_k\dot\beta_{n+2}}\,\phi^{(k,n+2)}_{\alpha_1...\alpha_{k}\,\dot\beta_1...\dot\beta_{n+1}\dot\beta_{n+2}} \,,\qquad k\geq 1\,;
\end{equation} \begin{equation}\label{Eq-Dir-comp2-a}
\partial^{\,\dot\beta_n}_{\alpha_{k+1}}\,\phi^{(k,n)}_{\alpha_1...\alpha_{k}\,\dot\beta_1...\dot\beta_n}=
iem\,x^{\alpha_{k+2}\dot\beta_{n}}\,\phi^{(k+2,n)}_{\alpha_1...\alpha_{k+1}\alpha_{k+2}\,\dot\beta_1...\dot\beta_{n}} \,,\qquad n\geq 1\,, \end{equation}
which are the Maxwell-invariant generalizations of Dirac-Pauli-Fierz equations. \item The last constraint  (\ref{constr-x3}) takes in first-quantized
theory the following form \begin{equation}\label{q-constr-x3-a}
\left[-\partial_{\alpha\dot\alpha}\partial^{\alpha\dot\alpha}-2i\partial^{\alpha\dot\alpha}\partial_{\alpha}\bar\partial_{\dot\alpha}+ 2iem
\left(x^{\beta}_{\dot\alpha}\,\partial^{\alpha\dot\alpha}\partial_{\alpha}\partial_{\beta}+
x_{\alpha}^{\dot\beta}\,\partial^{\alpha\dot\alpha}\bar\partial_{\dot\alpha}\bar\partial_{\dot\beta}\right)
-2e^2m^2\left(\partial_{\alpha}x^{\alpha\dot\alpha}\bar\partial_{\dot\alpha}\right)^2\,\right]\Phi= 0\,. \end{equation} However, from
(\ref{q-costr-x1-a}), (\ref{q-costr-z1-a}) and  (\ref{Eq-Psi-a}) follows the equations \begin{equation}\label{eq-constr-dop-a}
ix^{\beta}_{\dot\alpha}\,\partial^{\alpha\dot\alpha}\partial_{\alpha}\partial_{\beta}\,\Phi=
ix_{\alpha}^{\dot\beta}\,\partial^{\alpha\dot\alpha}\bar\partial_{\dot\alpha}\bar\partial_{\dot\beta}\,\Phi=
em\,\left(\partial_{\alpha}x^{\alpha\dot\alpha}\bar\partial_{\dot\alpha}\right)^2\,\Phi\,. \end{equation} Using the relations (\ref{eq-constr-dop-a}) and
(\ref{Tr-Psi-a}) we can reduce the constraint (\ref{q-constr-x3-a}) to the following form \begin{equation}\label{q-constr-x3-Psi-a}
\left[-\partial_{\alpha\dot\alpha}\partial^{\alpha\dot\alpha}
+2e^2m^2\left(\partial_{\alpha}x^{\alpha\dot\alpha}\bar\partial_{\dot\alpha}\right)^2\,\right]\Phi= 0\,. \end{equation} The equation
(\ref{q-constr-x3-Psi-a}) provides the following infinite set of component field equations \begin{equation}\label{Eq-KG-comp-a}
\partial^{\gamma\dot\gamma}\partial_{\gamma\dot\gamma}\,\phi^{(k,n)}_{\alpha_1...\alpha_{k}\,\dot\beta_1...\dot\beta_n}=
2e^2m^2\,x^{\alpha_{k+1}\dot\beta_{n+1}}\,x^{\alpha_{k+2}\dot\beta_{n+2}}\,
\phi^{(k+2,n+2)}_{\alpha_1...\alpha_{k}\alpha_{k+1}\alpha_{k+2}\,\dot\beta_1...\dot\beta_{n}\dot\beta_{n+1}\dot\beta_{n+2}} \,, \end{equation} which are
Maxwell-invariant generalization of massless Klein-Gordon equations. One can demonstrate that the Maxwell-Klein-Gordon equations (\ref{Eq-KG-comp-a}) for
all component fields $\phi^{(k,n)}$, except $\phi^{(0,0)}$, can be derived from the Maxwell-Dirac equations (\ref{Eq-Dir-comp1-a}), (\ref{Eq-Dir-comp2-a}).
Single additional equation for the component fields following from  (\ref{Eq-KG-comp-a}) is the Maxwell-Klein-Gordon equation for the scalar field
$\phi^{(0,0)}$: \begin{equation}\label{Eq-KG-comp-0-a} \partial^{\gamma\dot\gamma}\partial_{\gamma\dot\gamma}\,\phi^{(0,0)}=
2e^2m^2\,x^{\alpha_{1}\dot\beta_{1}}\,x^{\alpha_{2}\dot\beta_{2}}\, \phi^{(2,2)}_{\alpha_1\alpha_{2}\,\dot\beta_1\dot\beta_{2}} \,. \end{equation}
\end{itemize}

In order to specify in our model the irreducible Maxwell-HS multiplets let us represent down the Maxwell algebra generators  (\ref{M-gen-cl}) in terms of
realizations (\ref{op-real1-a}), (\ref{op-real2-a}). We obtain that the infinitesimal Maxwell symmetry transformations are the following
\begin{equation}\label{M-gen-qu-a} \begin{array}{rcl} P_{\alpha\dot\beta}&:\quad&\delta_P\phi^{(k,n)}_{\,\cdots}=
a^{\gamma\dot\gamma}\partial_{\gamma\dot\gamma}\phi^{(k,n)}_{\,\cdots}+
iem\,a^{\alpha\dot\gamma}\partial^{\beta}_{\dot\gamma}\phi^{(k+2,n)}_{\,\alpha\beta\,\cdots} +
iem\,a^{\dot\alpha\gamma}\partial^{\dot\beta}_{\gamma}\phi^{(k,n+2)}_{\,\cdots\, \dot\alpha\dot\beta}\,,\\[5pt]
Z_{\alpha\beta}&:\quad&\delta_Z\phi^{(k,n)}_{\,\cdots}= -i em\,b^{\alpha\beta}\phi^{(k+2,n)}_{\,\alpha\beta\,\cdots} \,,\\[5pt] \bar
Z_{\dot\alpha\dot\beta}&:\quad&\delta_{\bar Z}\phi^{(k,n)}_{\,\cdots}= -i em\,\bar b^{\dot\alpha\dot\beta}\phi^{(k,n+2)}_{\,\cdots\,
\dot\alpha\dot\beta}\,,\\[5pt] M_{\alpha\beta},\bar M_{\dot\alpha\dot\beta}&:\quad&\delta_M\phi^{(k,n)}_{\,\alpha\,\cdots\,\dot\alpha}=
-\left(\ell^{\beta\gamma}x^{\dot\gamma}_{\beta}+
\bar\ell^{\dot\beta\dot\gamma}x^{\gamma}_{\dot\beta}\right)\partial_{\gamma\dot\gamma}\phi^{(k,n)}_{\,\alpha\,\cdots\,\dot\alpha}+ k\,
\ell^{\beta}_{\alpha}\phi^{(k,n)}_{\,\beta\,\cdots\,\dot\alpha} + n\, \bar\ell^{\dot\beta}_{\dot\alpha}\phi^{(k,n)}_{\,\alpha\,\cdots\,\dot\beta}\,.
\end{array} \end{equation}

From the relations  (\ref{M-gen-qu-a}) follows that one can construct the infinite-dimensional Maxwell-HS multiplets with minimal Lorentz spin
$(\frac{k}{2},\frac{n}{2})$ described by the field $\phi^{(k,n)}(x)$ if we supplement the infinite chain of component fields $\phi^{(k+2p,n+2r)}(x)$ where
$p=0,1,2,...$, $r=0,1,2,...$. One can observe that \begin{description} \item[i)] There are two infinite sets (bosonic and fermionic) of
infinite-dimensional fields: with integer spins $s=\frac{1}{2}(k+n)+p+r$ ($k+n$ even) and with half-integer spins ($k+n$ odd). \item[ii)] The field
equations relating the components $\phi^{(k+2p,n+2r)}(x)$ are given by the Maxwell-Weyl equations (\ref{Eq-Dir-comp1-a}), (\ref{Eq-Dir-comp2-a}) and
supplementary equations (\ref{Eq-comp-a}); only for spin-zero field $\phi^{(0,0)}$ we should supplement the Maxwell-Klein-Gordon equations
(\ref{Eq-KG-comp-0-a}). In particular, the maximal bosonic Maxwell HS multiplet with scalar field $\phi^{(0,0)}(x)$ will contain the link between all
components with even Lorents spins $(j_1,j_2)=(p,r)$, and there are two maximal fermionic multiplets: chiral, with Maxwell-Weyl field
$\phi^{(1,0)}_{\alpha}(x)$, and the antichiral one , with Maxwell-Weyl field $\phi^{(0,1)}_{\dot\alpha}(x)$ (in Majorana case they are related by the
relation $\phi^{(0,1)}_{\dot\alpha}(x)=(\phi^{(1,0)}_{\alpha}(x))^\dagger$). \item[iii)] The equations (\ref{Eq-comp-a}), (\ref{Eq-Dir-comp1-a}) and
(\ref{Eq-Dir-comp2-a}) link the fields with different spins $(k,n)$. For $k\geq 1$ and $n\geq 1$ the table of fields $\phi^{(k,n)}(x)$ can be decomposed
into triplet of spins with the closed diagram \begin{equation}\label{d-close} (k,n) \quad \stackrel{(3.18)}{\dashleftarrow\dashrightarrow} \quad (k,n+2)
\quad \stackrel{(3.16)}{\dashleftarrow\dashrightarrow} \quad  (k+2,n) \quad \stackrel{(3.19)}{\dashleftarrow\dashrightarrow} \quad  (k,n) \end{equation}
where over the double arrows we indicate the field equation which connects different spins. For the fields with spins $(k,0)$ and $(0,n)$ one can not
construct closed sequence of links, however one can relate all the bosonic ($k+n$ even) or fermionic ($k+n$ odd) fields of a such type (they correspond to
self-dual curvatures of higher spin gauge fields). One can compose the following two sequences \begin{equation}\label{d-unclose} \begin{array}{lll} &&
(k,0) \quad \stackrel{(3.18)}{\dashleftarrow\dashrightarrow} \quad (k,2) \quad \stackrel{(3.16)}{\dashleftarrow\dashrightarrow} \quad  (k+2,0)  \\[5pt] &&
(0,n) \quad \stackrel{(3.19)}{\dashleftarrow\dashrightarrow} \quad (2,n) \quad \stackrel{(3.16)}{\dashleftarrow\dashrightarrow} \quad  (0,n+2) \end{array}
\end{equation} \end{description}

We stress that the coupling between different spins described in (\ref{d-close}), (\ref{d-unclose}) by double arrows is
proportional to $em$. In particular if $e\to 0$ the fields with different spins are decoupled, and the equations (\ref{Eq-Dir-comp1-a}),
(\ref{Eq-Dir-comp2-a}) describe the Dirac-Pauli-Fierz equations (\ref{Dir-m0-a}) for free HS fields; the limit $e\to 0$ of  (\ref{Eq-KG-comp-0-a}) describes free
Klein-Gordon equation for spinless field.

We emphasize that performing the limit $e\to 0 $ on the level of the action with Lagrangian (\ref{Maxwell-Lagr})
and subsequent quantization provides only subset of free fields described by equations (\ref{Dir-m0-a}).
In such free HS model the spectrum contains only the scalar field $\phi^{(0,0)}(x)$,
the spin-tensor fields with undotted indices $\phi^{(k,0)}_{\alpha_1...\alpha_k}(x)$ ($k>0$) and spin-tensor field
with dotted indices $\phi^{(0,n)}_{\dot\alpha_1...\dot\alpha_n}(x)$ ($n>0$).
This is due to the property of the action (\ref{Maxwell-Lagr}) in the limit $e\to 0 $, in which all ten constraints in the system become of the first class,
in particular all the constraints (\ref{costr-x}). Corresponding unfolded equations
$\left( \partial_{\alpha\dot\beta}+ i  \partial_{\alpha}\bar\partial_{\dot\beta}\right)\Phi=0$ yield besides the equations
(\ref{Dir-m0-a}) and the Klein-Gordon equation for scalar field $\phi^{(0,0)}(x)$ also the constraints which express `mixed' fields
$\phi^{(k>0,n>0)}_{\alpha_1...\alpha_k\,\dot\beta_1...\dot\beta_n}(x)$ in terms of space-time
derivatives of the fields $\phi^{(0,0)}(x)$, $\phi^{(k>0,0)}_{\alpha_1...\alpha_k}(x)$, $\phi^{(0,n>0)}_{\dot\alpha_1...\dot\alpha_n}(x)$
(see details in \cite{BLS,Vas}).

If $e\neq 0$ it was shown in Sect.\,3.1 that only eight out of ten constraints (\ref{costr-x})-(\ref{costr-by})
are first class and one can introduce the fields $\phi^{(k,n)}$ for $k\neq 0$, $n\neq 0$ as independent.
Such fields enter together with the ``unfolded'' field $\phi^{(k,0)}_{\alpha_1...\alpha_k}(x)$ into
the infinite-dimensional multiplet $\phi^{(k+2p,2r)}$ which describes an infinite-dimensional realization
(see (\ref{M-gen-qu-a})) of $D=4$ Maxwell algebra.

Finally we add that the appearance of multiplicative massive parameter $m$ can not be avoided; it could be traced to nonvanishing
dimensionalities of fourmomenta $P_\mu$ and the space-time coordinates. In principle one can only rearrange the dimensionality $[Z_{\mu\nu}]$ of the
generators $Z_{\mu\nu}$, but as we show in the Appendix, if we consistently remove $m$ in the Lagrangian (\ref{Maxwell-Lagr}) by assuming that
$[Z_{\mu\nu}]=1$, we will be forced to introduce massive parameter $M$ into the Maxwell algebra relations (\ref{M-alg}) as well as in Maurer-Cartan one
forms (\ref{M-om}).

\subsection{HS field equations in dual picture for tensorial coordinates}

We can also consider the Schr\"{o}dinger representation in which the wave function is related by Fourier transform in the tensorial sector
\begin{equation}\label{WF} \tilde\Psi =\tilde\Psi(x^{\alpha\dot\beta},f_{\alpha\beta}, \bar f_{\dot\alpha\dot\beta}, y^\alpha,\bar y^{\dot\alpha})\,.
\end{equation} The generalized dual momenta variables are realized by the following partial derivatives \begin{equation}\label{op-real1}
z^{\alpha\beta}=i\frac{\partial\,\,}{\partial f_{\alpha\beta}}\,,\qquad \bar z^{\dot\alpha\dot\beta}=i\frac{\partial\,\,}{\partial \bar
f_{\dot\alpha\dot\beta}}\,. \end{equation} The Maxwell-covariant quantum momenta $D_{\alpha\dot\beta}$ satisfy the relation
\begin{equation}\label{d-costr-com} \left[D_{\alpha\dot\alpha}, D_{\beta\dot\beta}\right]= 2i\,e \left(\epsilon_{\dot\alpha\dot\beta} f_{\alpha\beta}+
\epsilon_{\alpha\beta}\bar f_{\dot\alpha\dot\beta}\right) \end{equation} and describe the Maxwell-covariant derivative in Schr\"{o}dinger realization
(\ref{op-real1}). The field equations are defined by the quantum counterpart of first class constraints  (\ref{costr-x1})-(\ref{costr-y1}):
\begin{eqnarray}\label{q-costr-x1} i D_{\alpha\dot\beta}\bar\partial^{\dot\beta}\,\tilde\Psi&=&\left(\partial_{\alpha\dot\beta}+ i
\,e\,f_{\alpha\gamma}\,x_{\dot\beta}^{\gamma}\right)\bar\partial^{\dot\beta}\,\tilde\Psi= 0\, ,\\[5pt] \label{q-costr-x1a} i
D_{\beta\dot\alpha}\partial^{\beta}\,\tilde\Psi&=&\left(\partial_{\beta\dot\alpha}+ i \,e\,\bar f_{\dot\alpha\dot\gamma} \,x_{\beta}^{\dot\gamma}
\right)\partial^{\beta}\, \tilde\Psi= 0\, , \end{eqnarray} \begin{equation}\label{q-costr-y1} \Big(f_{\alpha\beta}+
 m\partial_{\alpha} \partial_{\beta}\Big)\tilde\Psi=0\,,\qquad\qquad
\Big(\bar f_{\dot\alpha\dot\beta}+ m \bar\partial_{\dot\alpha}\bar\partial_{\dot\beta}\Big)\tilde\Psi=0\,. \end{equation}

Let us consider the wave functions with polynomial dependence with respect to $y\equiv y^{\alpha}$ and $\bar y\equiv \bar y^{\dot\alpha}$
\begin{equation}\label{WF-1} \tilde\Psi(x,f, \bar f, y,\bar y)= \sum_{k,n=0}^{\infty}\frac{1}{k!\,n!}\,y^{\alpha_1}...y^{\alpha_k}\, \bar
y^{\dot\beta_1}...\bar y^{\dot\beta_n} \,\tilde\Psi^{(k,n)}_{\alpha_1...\alpha_k\,\dot\beta_1...\dot\beta_n}(x,f,\bar f)\,. \end{equation} Then the
constraints (\ref{q-costr-y1}) for the wave function (\ref{WF}) lead to expressing of all components $\tilde\Psi^{(k,n)}$, $k>1$, $n>1$ by
$\tilde\Psi^{(0,0)}$, $\tilde\Psi^{(1,0)}$, $\tilde\Psi^{(0,1)}$ and $\tilde\Psi^{(1,1)}$. For example, $$ \tilde\Psi^{(2,0)}_{\alpha\beta}= -
m^{-1}f_{\alpha\beta} \tilde\Psi^{(0,0)}\,,\qquad \tilde\Psi^{(0,2)}_{\dot\alpha\dot\beta}= - m^{-1}\bar f_{\dot\alpha\dot\beta} \tilde\Psi^{(0,0)}\,, $$
$$ \tilde\Psi^{(2,1)}_{\alpha\beta\dot\gamma}= - m^{-1}f_{\alpha\beta} \tilde\Phi^{(0,1)}_{\dot\gamma}\,,\qquad
\tilde\Psi^{(1,2)}_{\gamma\dot\alpha\dot\beta}= - m^{-1}\bar f_{\dot\alpha\dot\beta} \tilde\Psi^{(1,0)}_{\gamma}\,, $$ $$ \qquad \qquad
\tilde\Psi^{(3,1)}_{\alpha\beta\gamma\dot\gamma}= - m^{-1}f_{\alpha\beta} \tilde\Psi^{(1,1)}_{\gamma\dot\gamma}\,,\qquad
\tilde\Psi^{(1,3)}_{\gamma\dot\alpha\dot\beta\dot\gamma}= - m^{-1}\bar f_{\dot\alpha\dot\beta} \tilde\Psi^{(1,1)}_{\gamma\dot\gamma}\,,\qquad \mbox{etc}.
$$

Thus, if $m\neq 0$ the fields in the expansion  (\ref{WF-1}) are determined by ten-dimensional ``generalized spin zero'' field
\begin{equation}\label{g-spin0} \tilde\Psi^{(0,0)}(x,f,\bar f)\,, \end{equation} two ``generalized spin half'' fields \begin{equation}\label{g-spin12}
\tilde\Psi^{(1,0)}_{\alpha}(x,f,\bar f)\,,\qquad \tilde\Psi^{(0,1)}_{\dot\beta}(x,f,\bar f) \end{equation} and ``generalized spin one'' field
\begin{equation}\label{g-spin1} \tilde\Psi^{(1,1)}_{\alpha\dot\beta}(x,f,\bar f)\,. \end{equation} The constraints (\ref{q-costr-y1}) provide the following
constraints for the fields (\ref{g-spin0}), (\ref{g-spin12}), (\ref{g-spin1}): \begin{equation}\label{con-spin0}
f^{\alpha\beta}f_{\alpha\beta}\tilde\Psi^{(0,0)}(x,f,\bar f)=0\,,\qquad \bar f^{\dot\alpha\dot\beta}\bar f_{\dot\alpha\dot\beta}\tilde\Psi^{(0,0)}(x,f,\bar
f)=0\,, \end{equation} \begin{equation}\label{con-spin12} f_{\alpha}{}^{\beta}\tilde\Psi^{(1,0)}_{\beta}(x,f,\bar f)=0\,,\qquad \bar
f_{\dot\alpha}{}^{\dot\beta}\tilde\Psi^{(0,1)}_{\dot\beta}(x,f,\bar f)=0\,, \end{equation} \begin{equation}\label{con-spin1}
f_{\alpha}{}^{\beta}\tilde\Psi^{(1,1)}_{\beta\dot\beta}(x,f,\bar f)=0\,,\qquad \bar
f_{\dot\alpha}{}^{\dot\beta}\tilde\Psi^{(1,1)}_{\beta\dot\beta}(x,f,\bar f)=0\,. \end{equation}

{}From the equations (\ref{q-costr-x1}), (\ref{q-costr-x1a}) for the unconstrained component (\ref{g-spin12}), (\ref{g-spin1}) one obtain the relations
\begin{equation}\label{Dir-12} \left(\partial^{\alpha\dot\beta}- i
\,e\,f^{\alpha\gamma}\,x^{\dot\beta}_{\gamma}\right)\tilde\Psi^{(0,1)}_{\dot\beta}(x,f,\bar f)= 0\, ,\qquad \left(\partial^{\beta\dot\alpha}- i \,e\,\bar
f^{\dot\alpha\dot\gamma} \,x^{\beta}_{\dot\gamma} \right)\tilde\Psi^{(1,0)}_{\beta}(x,f,\bar f)= 0\, , \end{equation} \begin{equation}\label{Dir-1}
\left(\partial^{\alpha\dot\beta}- i  \,e\,f^{\alpha\gamma}\,x^{\dot\beta}_{\gamma}\right)\tilde\Psi^{(1,1)}_{\beta\dot\beta}(x,f,\bar f)= 0\, ,\qquad
\left(\partial^{\beta\dot\alpha}- i \,e\,\bar f^{\dot\alpha\dot\gamma} \,x^{\beta}_{\dot\gamma} \right)\tilde\Psi^{(1,1)}_{\beta\dot\beta}(x,f,\bar f)= 0\,
, \end{equation} which have the form of the Dirac equations in a constant electromagnetic field, with electromagnetic potential
$\mathscr{A}_{\mu}=f_{\mu\nu}x^\nu$. Contrary to the standard approach for Dirac spin-half field, the wave functions in (\ref{Dir-12}) depend also on
continuous electromagnetic field strength coordinates $f_{\alpha\beta}$, $\bar f_{\dot\alpha\dot\beta}$. This additional dependence of the wave functions
can not be generated only by minimal coupling of the external electromagnetic field. We do not see yet the relationship of our description of interacting
HS fields to the approaches proposed in recent papers \cite{PRSag,BuchSZ}, however it would be interesting to find such a link.

Generalized spin-zero field $\tilde\Psi^{(0,0)}(x,f,\bar f)$ is described by generalized Klein-Gordon equation, which follows from the constraint
\begin{equation}\label{KG-gen} \hat T^{\alpha\dot\alpha}\hat T_{\alpha\dot\alpha}\,\tilde\Psi\approx 0\,. \end{equation} Taking into account that
\begin{equation}\label{KG-op} \begin{array}{rcl} \hat T^{\alpha\dot\alpha}\hat T_{\alpha\dot\alpha}&=&
-\partial^{\alpha\dot\alpha}\partial_{\alpha\dot\alpha} +2ie\left(f^{\alpha\beta}x^{\dot\alpha}_{\beta} +\bar
f^{\dot\alpha\dot\beta}x^{\alpha}_{\dot\beta}\right)\partial_{\alpha\dot\alpha}\\[6pt] && +{\textstyle\frac12}\,e^2\left(f^{\alpha\beta}f_{\alpha\beta} +
\bar f^{\dot\alpha\dot\beta}\bar f_{\dot\alpha\dot\beta}\right)x^{\gamma\dot\gamma}x_{\gamma\dot\gamma} -2e^2f_{\alpha\beta}\bar
f_{\dot\alpha\dot\beta}x^{\beta\dot\alpha}x^{\alpha\dot\beta}\\[6pt] && -2\left(i\partial^{\alpha\dot\alpha}+ef^{\alpha\beta}x^{\dot\alpha}_{\beta} + e\bar
f^{\dot\alpha\dot\beta}x^{\alpha}_{\dot\beta}\right)\partial_{\alpha}\bar\partial_{\dot\alpha}\,, \end{array} \end{equation} we obtain the generalized
Klein-Gordon equation for spin-zero field \begin{equation}\label{KG-0} \left[ -\Box +2ie\left(f^{\alpha\beta}x^{\dot\alpha}_{\beta} +\bar
f^{\dot\alpha\dot\beta}x^{\alpha}_{\dot\beta}\right)\partial_{\alpha\dot\alpha} +{\textstyle\frac12}\,e^2\Big( f^2+\bar f^2 \Big)x^{2}
-2e^2f_{\alpha\beta}\bar f_{\dot\alpha\dot\beta}x^{\beta\dot\alpha}x^{\alpha\dot\beta}\right]\tilde\Psi^{(0,0)}=0\,, \end{equation} where
$\Box:=\partial^{\alpha\dot\alpha}\partial_{\alpha\dot\alpha}$, $x^2:=x^{\alpha\dot\alpha}x_{\alpha\dot\alpha}$, $f^2:=f^{\alpha\beta}f_{\alpha\beta}$,
$\bar f^2:=\bar f^{\dot\alpha\dot\beta}\bar f_{\dot\alpha\dot\beta}$. It should be emphasized that due to the equations (\ref{Dir-1}) and the constraints
(\ref{con-spin1}) the last term in the operator (\ref{KG-op}) does not contribute to the equation (\ref{KG-0}) and we obtain finally the Klein-Gordon
equation in constant EM field.

The solutions of the equations  (\ref{con-spin0})-(\ref{con-spin1}) can be represented as the Fourier transforms \begin{eqnarray}
\tilde\Psi^{(0,0)}(x,f,\bar f)&=&\int d{}^{\,6}\!z\, e^{-i(fz+\bar f \bar z)} \,\Psi^{(0,0)}(x,z,\bar z) \,,\label{ser0}\\
\tilde\Psi^{(1,0)}_{\gamma}(x,f,\bar f)&=&\int d{}^{\,6}\!z\, e^{-i(fz+\bar f \bar z)} \,\Psi^{(1,0)}_{\gamma}(x,z,\bar z) \,,\label{ser12a}\\
\tilde\Psi^{(0,1)}_{\dot\gamma}(x,f,\bar f)&=&\int d{}^{\,6}\!z\, e^{-i(fz+\bar f \bar z)} \,\Psi^{(0,1)}_{\dot\gamma}(x,z,\bar z)\,,\label{ser12b}\\
\tilde\Psi^{(1,1)}_{\gamma\dot\gamma}(x,f,\bar f)&=&\int d{}^{\,6}\!z\, e^{-i(fz+\bar f \bar z)} \,\Psi^{(1,1)}_{\gamma\dot\gamma}(x,z,\bar z)
\,,\label{ser1} \end{eqnarray} where \begin{eqnarray} \Psi^{(0,0)}(x,z,\bar z)&=&
\sum_{k,n=0}^{\infty}\frac{1}{k!n!}\,z^{\alpha_1\beta_1}...z^{\alpha_k\beta_k}\, \bar z^{\dot\alpha_1\dot\beta_1}...\bar z^{\dot\alpha_n\dot\beta_n}
\,\phi^{(2k,2n)}_{(\alpha_1\beta_1...\alpha_k\beta_k)\,(\dot\alpha_1\dot\beta_1...\dot\alpha_n\dot\beta_n)}(x)\,,\label{com-ser0}\\
\Psi^{(1,0)}_{\gamma}(x,z,\bar z)&=& \sum_{k,n=0}^{\infty}\frac{1}{k!n!}\,z^{\alpha_1\beta_1}...z^{\alpha_k\beta_k}\, \bar
z^{\dot\alpha_1\dot\beta_1}...\bar z^{\dot\alpha_n\dot\beta_n}
\,\phi^{(2k+1,2n)}_{(\gamma\alpha_1\beta_1...\alpha_k\beta_k)\,(\dot\alpha_1\dot\beta_1...\dot\alpha_n\dot\beta_n)}(x)\,,\label{com-ser12a}\\
\Psi^{(0,1)}_{\dot\gamma}(x,z,\bar z)&=& \sum_{k,n=0}^{\infty}\frac{1}{k!n!}\,z^{\alpha_1\beta_1}...z^{\alpha_k\beta_k}\, \bar
z^{\dot\alpha_1\dot\beta_1}...\bar z^{\dot\alpha_n\dot\beta_n}
\,\phi^{(2k,2n+1)}_{(\alpha_1\beta_1...\alpha_k\beta_k)\,(\dot\gamma\dot\alpha_1\dot\beta_1...\dot\alpha_n\dot\beta_n)}(x)\,,\label{com-ser12b}\\
\Psi^{(1,1)}_{\gamma\dot\gamma}(x,z,\bar z)&=& \sum_{k,n=0}^{\infty}\frac{1}{k!n!}\,z^{\alpha_1\beta_1}...z^{\alpha_k\beta_k}\, \bar
z^{\dot\alpha_1\dot\beta_1}...\bar z^{\dot\alpha_n\dot\beta_n}
\,\phi^{(2k+1,2n+1)}_{(\gamma\alpha_1\beta_1...\alpha_k\beta_k)\,(\dot\gamma\dot\alpha_1\dot\beta_1...\dot\alpha_n\dot\beta_n)}(x)\label{com-ser1}
\end{eqnarray} have polynomial dependence on $z$, $\bar z$ and the component fields $\phi^{(k,n)}_{(\alpha...)\,(\dot\alpha...)}$ are symmetric with
respect to all undotted and dotted spinor indices, as the component fields considered in previous subsection. The fields (\ref{ser0})-(\ref{ser1}) have
complicated, nonpolynomial dependence on the variables $f$, $\bar f$. In conclusion, generalized Dirac equations (see (\ref{Dir-12})-(\ref{Dir-1})) and
Klein-Gordon equation (\ref{KG-0}) yield nontrivial, nonminimal coupling of these component fields to the constant EM field.

\subsection{The link between constant torsion in tensorial sector and \\ constant EM field background}

{}From the equations (\ref{Dir-12}), (\ref{Dir-1}) and (\ref{KG-0}) we see that for the link of different spins there is responsible EM coupling described
by nonvanishing parameter $e$. Indeed one can show that if the torsion in six tensorial dimensions of Maxwell space-time depends on the $D=4$ space-time
coordinates it can be reinterpreted as a coupling to an Abelian gauge potential. Let us introduce the following ``block-diagonal'' 10-bein
$E_{AB}{}^{CD}=\left( \delta_\alpha^\gamma \delta_{\dot\beta}^{\dot\delta}, E_{\alpha\dot\beta}{}^{\gamma\delta},
E_{\alpha\dot\beta}{}^{\dot\gamma\dot\delta}\right)$ in the tensorial space $\left( x^{\alpha\dot\beta}, z^{\alpha\beta}, \bar
z^{\dot\alpha\dot\beta}\right)$ \begin{equation}\label{cov-1-N} \nabla_{\alpha\dot\beta}=\partial_{\alpha\dot\beta} +
E_{\alpha\dot\beta}{}^{\gamma\delta}(x)\nabla_{\gamma\delta}+ E_{\alpha\dot\beta}{}^{\dot\gamma\dot\delta}(x)\nabla_{\dot\gamma\dot\delta}\,,
\end{equation} \begin{equation}\label{cov-2-N} \nabla_{\alpha\beta}=\partial_{\alpha\beta}\,,\qquad
\nabla_{\dot\alpha\dot\beta}=\partial_{\dot\alpha\dot\beta}\,. \end{equation} If we consider the plane wave solutions in additional dimensions, one can
replace (see (\ref{op-real1-a})) the derivatives  (\ref{cov-2-N}) by constant tensors $f_{\alpha\beta}$, $\bar f_{\dot\alpha\dot\beta}$ describing
additional tensorial momenta. In such a case the derivative  (\ref{cov-1-N}) can be written as the Abelian gauge-covariant derivative
\begin{equation}\label{cov-3-N} \nabla_{\alpha\dot\beta}=\partial_{\alpha\dot\beta} + \mathscr{A}_{\alpha\dot\beta}(x)\,, \end{equation} where
$\mathscr{A}_{\alpha\dot\beta}(x)=E_{\alpha\dot\beta}{}^{\gamma\delta}(x)f_{\gamma\delta}+ E_{\alpha\dot\beta}{}^{\dot\gamma\dot\delta}(x)\bar
f_{\dot\gamma\dot\delta}$. In Maxwell tensorial space additional tensorial coordinates are twisted by a constant torsion, the functions
$E_{\alpha\dot\beta}{}^{\gamma\delta}(x)$ and $E_{\alpha\dot\beta}{}^{\dot\gamma\dot\delta}(x)$ are linear in $x$, and we obtain in (\ref{cov-3-N}) the
Abelian gauge field four-potential $\mathscr{A}_{\alpha\dot\beta}$ describing constant electromagnetic field strength ($\bar
f_{\dot\alpha\dot\beta}=(f_{\alpha\beta})^\dagger$) \begin{equation}\label{const-A} \mathscr{A}_{\alpha\dot\beta}=
f_{\alpha}^{\,\gamma}x_{\gamma\dot\beta}+\bar f_{\dot\beta}^{\,\dot\gamma}x_{\alpha\dot\gamma}\,. \end{equation} It can be added that the translations
$x_{\alpha\dot\beta}\to x_{\alpha\dot\beta} +a_{\alpha\dot\beta}$ modify (\ref{const-A}) by a constant term, which can be however compensated by the
Abelian gauge transformation of $\mathscr{A}_{\alpha\dot\beta}$, what leads to the translational invariance of covariant derivative
(\ref{cov-3-N}).\footnote{We thank E.\,Ivanov for this observation.}

\setcounter{equation}{0} \section{Final remarks}

In this paper we introduced in ten-dimensional tensorial space the Maxwell-invariant spinorial particle model with auxiliary spinor variables, which after
its first quantization provides the generalization of massless conformal HS fields and define the infinite-dimensional Maxwell-HS field multiplets. These
multiplets of Maxwell HS fields should be further studied, in particular their relation with free massless conformal HS fields and the possibility of their
derivation from an action principle. New multiplets describe particular field-theoretic realizations of Maxwell algebra with vanishing four Casimirs
$C^{Max}_1$, $C^{Max}_2$, $C^{Max}_3$, $C^{Max}_4$ (see (\ref{Cas-fix})). We recall that the mass-shall condition $C^{Max}_1=0$, due to the presence in
$C^{Max}_1$ of the term proportional to relativistic angular momentum ($M_{\alpha\beta}$, $\bar M_{\dot\alpha\dot\beta}$), can be possibly linked with the
description of spin-dependent mass spectrum for Regge trajectories.

The infinite-dimensional sets of Maxwell-HS field which define the realization $T^{(k,n)}$ of Maxwell algebra is para\-met\-rized by two natural numbers
$k=0,1,2,...$, $n=0,1,2,...$. These numbers describe the minimal value of Lorentz spins ($j_1{=}\frac{k}{2},j_2{=}\frac{n}{2}$) in the multiplet of fields
$\phi^{(k+2p,n+2r)}_{\alpha_1...\alpha_{k+2p}\,\dot\beta_1...\dot\beta_{n+2r}}(x) \in T^{(k,n)}$, where $p=0,1,2,...$, $r=0,1,2,...$. It follows that the
representation $T^{(\tilde k,\tilde n)}$ if $\tilde k>k$, $\tilde n>n$ respectively for even and for odd values of ${( k, n)}$, ${(\tilde k,\tilde n)}$, is
included in the bosonic (fermionic) representation $T^{(k,n)}$, i.e.   $T^{(\tilde k,\tilde n)}\subset T^{(k,n)}$. The largest ``master'' bosonic
realization, with component fields having all integer spins $s=\frac12(j_1+j_2)$, is given by $T^{(0,0)}$; two maximal chiral and antichiral fermionic
realizations, with half-integer spins, are provided by $T^{(1,0)}$ and $T^{(0,1)}$.

By considering the particle model on extended space-time ($x_{\alpha\dot\beta}$, $z_{\alpha\beta}$, $\bar z_{\dot\alpha\dot\beta}$, $\lambda_{\alpha}$,
$\bar\lambda_{\dot\alpha}$) and its quantization we have shown how the field-theoretic realizations of Maxwell algebra in such ten-dimensional tensorial
space can be used for the introduction of infinite-dimensional $D=4$ Maxwell-HS field multiplets with coupled spin components. The Maxwell-covariant
description of $D=4$ Maxwell-HS fields requires the presence of space-time-dependent coupling terms between different spin fields which can be also
interpreted (see (\ref{Dir-12}), (\ref{Dir-1})) as following from the electromagnetic covariantization of space-time derivatives in the presence of
constant EM background field strength.

In conclusion we would like to comment that despite the remaining open questions we hope that our model provides a step in understanding of
the theory of interacting HS fields.

\section*{Acknowledgements}

\noindent The authors would like to thank J.\,Buchbinder, E.\,Ivanov, P.\,Kosinski  and D.\,Sorokin for valuable remarks. We acknowledge a support from the
grant of the Bogoliubov-Infeld Programme, the grant of the INFN-BLTP JINR collaboration and RFBR grants 09-01-93107, 11-02-90445, 12-02-00517 (S.F.), as
well as from the Polish Ministry of Science and Higher Educations grant No.~N202331139 and Polish National Center of Science (NCN) research project
No.~2011/01/ST2/03354 (J.L.). S.F. thanks the members of the Institute of Theoretical Physics at Wroclaw University and Dipartimento di Fisica ``Galileo
Galilei'' at Universit\'{a} degli Studi di Padova for the warm hospitality at different stages of this study. J.L. thanks CERN Theory Division for the
hospitality at the final stage of this work.

\renewcommand\theequation{A.\arabic{equation}} \setcounter{equation}0 \section*{Appendix:   Different dimensionalities of tensorial coordinates and some
consequences}

We can arbitrarily change the mass dimensionality of the generators ; consequently the relation (\ref{M-alg}) is replaced by relation (\ref{M-alg-m}).
One gets in rescaled case
\begin{equation}\label{dim-ch}
[Z_{\alpha\beta}]=[\bar Z_{\dot\alpha\dot\beta}]=2-\xi\,,
\end{equation}
what implies
\begin{equation}\label{dim-z}
[z^{\alpha\beta}]=[\bar z^{\dot\alpha\dot\beta}]=
[\omega_{(Z)}^{\alpha\beta}]=[\bar\omega_{(Z)}^{\dot\alpha\dot\beta}]=-2+\xi\,.
\end{equation}

The parameters introduced in (\ref{M-alg-m}) will modify the formulae (\ref{M-om}) as follows \begin{equation}\label{M-om-m} \omega^{\alpha\beta}=   d
z^{\alpha\beta} + eM^{\,\xi}x^{(\alpha\dot\gamma}d x^{\beta)}_{\dot\gamma}\,,\qquad \bar\omega^{\dot\alpha\dot\beta}= d\bar z^{\dot\alpha\dot\beta}+
eM^{\,\xi}x^{\gamma(\dot\alpha}d x^{\dot\beta)}_{\gamma}\,. \end{equation} After inserting relations (\ref{M-om-m}) in (\ref{Maxwell-Lagr}) it follows from
(\ref{dim-z}) that dimensionless action requires the replacement of $m$ by $m^{1-\xi}$. Subsequently we obtain the following modified constraints
(\ref{costr-x})-(\ref{costr-by}) \begin{eqnarray}\label{costr-x-m} T_{\alpha\dot\beta}&=&p_{\alpha\dot\beta}-\lambda_{\alpha}\bar\lambda_{\dot\beta}+
\tilde em\Big(\lambda_{\alpha}\lambda_{\gamma}x^\gamma_{\dot\beta}+\bar\lambda_{\dot\beta}\bar\lambda_{\dot\gamma}x^{\dot\gamma}_{\alpha}\Big)\approx 0\,
,\\[5pt] T_{\alpha\beta}&=&f_{\alpha\beta}-m^{1-\xi}\lambda_{\alpha}\lambda_{\beta}\approx 0\,,\label{costr-y-m}\\[5pt] \bar T_{\dot\alpha\dot\beta}&=&\bar
f_{\dot\alpha\dot\beta}-m^{1-\xi}\bar\lambda_{\dot\alpha}\bar\lambda_{\dot\beta}\approx 0\,,\label{costr-by-m} \end{eqnarray} where the  rescaled
dimensionless coupling constant $\tilde e$ is given by the formula \begin{equation}\label{til-e} \tilde e= \left(\frac{M}{m}\right)^{\xi}e\,.
\end{equation} One can check that the constraints (\ref{costr-x-m}) lead to the basic field equations (\ref{Eq-Dir-comp1-a})-(\ref{Eq-Dir-comp2-a}) and
(\ref{Eq-KG-comp-a}) with the replacement $m\to \tilde e m$. One can also reduce the number of parameters by putting in (\ref{Maxwell-Lagr}) $m=M$. In such
a case $\tilde e=e$ and the theory contains only one geometric mass parameter, determined by the structure constant of the of Maxwell algebra.

We shall make the following comments: \begin{description} \item[i)] If $\xi=0$ and $e\neq 0$ the Maxwell algebra takes its original form \cite{Bac,Schr,B}
and describes the space-time geometry with constant electromagnetic background. We obtain the relations from Sect.\,1--3 with terms in the equations which
couple different spins by terms proportional to $em$. \item[ii)] If $\xi=1$ we have the relation $\tilde e m=eM$. The equations (\ref{costr-y-m}),
(\ref{costr-by-m}) are the same as in the particle model in \cite{BL,BLS,Vas,PST,BBAST,Vas1} providing massless HS free fields, however the eq.
(\ref{costr-x-m}) is modified due to the presence of generalized Maxwell space-time $(x_\mu,z_{\mu\nu})$ with torsion proportional to $eM$. The basic field
equations are obtained from (\ref{Eq-Dir-comp1-a}), (\ref{Eq-Dir-comp2-a}) and (\ref{Eq-KG-comp-a}) by the replacement $m\to M$, i.e. for $\xi=1$ the
dynamical mass parameter in our model disappears from the constraints (\ref{costr-x-m})-(\ref{costr-by-m}) and is transmuted into the geometric one.
\item[iii)] If $\xi=2$ in the Lagrangian  (\ref{Maxwell-Lagr}) the parameter $m$ is replaced by $m^{-1}$. Such choice of Maxwell algebra generators (for
$e=1$) was considered recently in \cite{AKLuk,Sor2,Durka} as the algebraic basis for the generalization of Einstein gravity but the corresponding particle
action  (\ref{Maxwell-act}) is singular in the limit $m\to 0$. \end{description}


\begin{thebibliography}{99}

\bibitem{BL} I.\,Bandos, J.\,Lukierski, {\it Tensorial Central Charges and New Superparticle Models with Fundamental Spinor Coordinates}, Mod. Phys. Lett.
    {\bf A14} (1999) 1257, {\tt arXiv:hep-th/9811022}.

\bibitem{BLS} I.\,Bandos, J.\,Lukierski, D.\,Sorokin, {\it Superparticle Models with Tensorial Central Charges}, Phys. Rev. {\bf D61} (2000) 045002, {\tt
    arXiv:hep-th/9904109}.

\bibitem{BLPS} I.\,Bandos, J.\,Lukierski, C.\,Preitschopf, D.\,Sorokin, {\it OSp supergroup manifolds, superparticles and supertwistors}, Phys. Rev. {\bf
    D61} (2000) 065009, {\tt arXiv:hep-th/9907113}.

\bibitem{Vas} M.A.\,Vasiliev, {\it Conformal Higher Spin Symmetries of 4d Massless Supermultiplets and $osp(L,2M)$ Invariant Equations in Generalized
    (Super)Space}, Phys. Rev. {\bf D66} (2002) 066006, {\tt arXiv:hep-th/0106149}.

\bibitem{PST} M.\,Plyushchay, D.\,Sorokin, M.\,Tsulaia, {\it Higher Spins from Tensorial Charges and $OSp(N|2n)$ Symmetry}, JHEP {\bf 0304} (2003) 013,
    {\tt arXiv:hep-th/0301067}; {\it GL Flatness of $OSp(1|2n)$ and Higher Spin Field Theory from Dynamics in Tensorial Spaces}, the Proceedings of the
    International Workshop ``Supersymmetries and Quantum Symmetries'' (SQS'03, Dubna, 24-29 July, 2003), {\tt arXiv:hep-th/0310297}.

\bibitem{BBAST} I.\,Bandos, X.\,Bekaert, J.A.\,de\,Azcarraga, D.\,Sorokin, M.\,Tsulaia,  {\it Dynamics of Higher Spin Fields and Tensorial Space}, JHEP
    {\bf 0505} (2005) 031, {\tt arXiv:hep-th/0501113}.

\bibitem{Vas1} M.A.\,Vasiliev, {\it Higher-Spin Theories and $Sp(2M)$ Invariant Space--Time}, the talk presented at the third Sakharov Conference in
    Physics, Moscow, 26-29 June 2002, {\tt arXiv:hep-th/0301235}; {\it On Conformal, SL(4,R) and Sp(8,R) Symmetries of 4d Massless Fields}, Nucl. Phys.
    {\bf B793} (2008) 469, {\tt arXiv:0707.1085 [hep-th]}.

\bibitem{Bac} H.\,Bacry, P.\,Combe, J.L.\,Richard, {\it Group-theoretical analysis of elementary particles in an external electromagnetic field. 1. The
    relativistic particle in a constant and uniform field.}, Nuovo Cim. {\bf A67} (1970) 267.

\bibitem{Schr} R.\,Schrader, {\it The Maxwell group and the quantum theory of particles in classical homogeneous electromagnetic fields}, Fortsch. Phys.
    {\bf 20} (1972) 701.

\bibitem{B} J.\,Beckers, V.\,Hussin, {\it Minimal Electromagnetic Coupling Schemes. II. Relativistic And Nonrelativistic Maxwell Groups}, J. Math. Phys.
    {\bf 24} (1983) 1295.


\bibitem{Sor} D.V.\,Soroka, V.A.\,Soroka, {\it Tensor extension of the Poincare algebra}, Phys. Lett. {\bf B607} (2005) 302, {\tt arXiv:hep-th/0410012}.

\bibitem{BG} S.\,Bonanos, J.\,Gomis, {\it A note on the Chevalley-Eilenberg Cohomology for the Galilei and Poincare Algebras}, J. Phys. {\bf A42} (2009)
    145206, {\tt arXiv:0808.2243 [hep-th]}; {\it Infinite Sequence of Poincare Group Extensions: Structure and Dynamics}, J. Phys. {\bf A43} (2010) 015201,
    {\tt arXiv:0812.4140 [hep-th]}.

\bibitem{BGKL} J.\,Gomis, K.\,Kamimura, J.\,Lukierski, {\it Deformations of Maxwell algebra and their Dynamical Realizations}, JHEP {\bf 0908} (2009) 039,
    {\tt arXiv:0906.4464 [hep-th]}.

\bibitem{AKLuk} J.A.\,de\,Azcarraga, K.\,Kamimura, J.\,Lukierski, {\it Generalized cosmological term from Maxwell symmetries}, Phys. Rev. {\bf D83} (2011)
    124036, {\tt arXiv:1012.4402 [hep-th]}.

\bibitem{Sor2} D.V.\,Soroka, V.A.\,Soroka, {\it Gauge semi-simple extension of the Poincare group}, {\tt arXiv:arXiv:1101.1591}.

\bibitem{Durka} R.\,Durka, J.\,Kowalski-Glikman, M.\,Szczachor, {\it Gauged AdS-Maxwell algebra and gravity}, {\tt arXiv:1107.4728 [hep-th]}.

\bibitem{FedLuk} S.\,Fedoruk, J.\,Lukierski, {\it New particle model in extended space-time and covariantization of planar Landau dynamics},
Phys. Lett.  {\bf B718} (2012) 646, {\tt arXiv:1207.5683 [hep-th]}.

\bibitem{Shir} T.\,Shirafuji, {\it Lagrangian Mechanics Of Massless Particles With Spin}, Prog. Theor. Phys. {\bf 70} (1983) 18.

\bibitem{FV87a} E.S.\,Fradkin, M.A.\,Vasiliev, {\it Cubic Interaction in Extended Theories of Massless Higher Spin Fields},
Nucl. Phys. {\bf B291} (1987) 141.

\bibitem{FV87b} E.S.\,Fradkin, M.A.\,Vasiliev, {\it On the Gravitational Interaction of Massless Higher Spin Fields},
Phys. Lett.  {\bf B189} (1987) 89.


\bibitem{BekBS} X.\,Bekaert, N.\,Boulanger, P.\,Sundell,
{\it How higher-spin gravity surpasses the spin two barrier: no-go theorems versus yes-go examples},
Rev. Mod. Phys.  {\bf 84} (2012) 987, {\tt arXiv:1007.0435 [hep-th]}.


\bibitem{Cur} T.\,Curtright, {\it Are there any superstrings in eleven-dimensions?}, Phys. Rev. Lett. {\bf 60} (1988) 393, Erratum-ibid. {\bf 60} (1988)
    1990.


\bibitem{AzTown} J.A.\,de\,Azcarraga, J.P.\,Gauntlett, J.M.\,Izquierdo, P.K.\,Townsend, {\it Topological Extensions of the Supersymmetry Algebra for
    Extended Objects}, Phys. Rev. Lett. {\bf 63} (1989) 2443.


\bibitem{Sezg} E.\,Sezgin, {\it The M algebra}, Phys. Lett. {\bf B392} (1997) 323, {\tt arXiv:hep-th/9609086}.

\bibitem{FZ} S.\,Fedoruk, V.G.\,Zima, {\it Massive Superparticle with Tensorial Central Charges}, Mod. Phys. Lett. {\bf A15} (2000) 2281, {\tt
    arXiv:hep-th/0009166}.

\bibitem{FI} S.\,Fedoruk, E.\,Ivanov, {\it Master Higher-Spin Particle}, Class. Quant. Grav. {\bf 23} (2006) 5195, {\tt arXiv:hep-th/0604111}.

\bibitem{SS-D} D.V.\,Soroka, V.A.\,Soroka, {\it Another approach to cosmological term problem}, talk at the International Workshop ``Supersymmetries and
    Quantum Symmetries'' (SQS'2011), Dubna, 18-23 July,  2011 ({\tt http://theor.jinr.ru/sqs/2011}).

\bibitem{PRSag} M.\,Porrati, R.\,Rahman, A.\,Sagnotti, {\it String Theory and The Velo-Zwanziger Problem},\\ Nucl. Phys.  {\bf B846} (2011) 250, {\tt
    arXiv:1011.6411 [hep-th]}.

\bibitem{BuchSZ} I.L.\,Buchbinder, T.V.\,Snegirev, Yu.M.\,Zinoviev, {\it Cubic interaction vertex of higher-spin fields with external electromagnetic
    field}, Nucl. Phys.  {\bf B864} (2012) 694, {\tt arXiv:1204.2341~[hep-th]}.


\end{thebibliography}
\end{document}